\documentclass[11pt]{article}

\usepackage[a4paper,margin=1in]{geometry}
\usepackage{setspace}
\usepackage[T1]{fontenc}
\usepackage[utf8]{inputenc}
\usepackage{lmodern}

\usepackage{amsmath}
\usepackage{amssymb}
\usepackage{amsfonts}
\usepackage{bm}

\usepackage{graphicx}
\usepackage{caption}
\usepackage{subcaption}
\usepackage{booktabs}
\usepackage{multirow}
\usepackage{array}
\usepackage{tabularx}
\usepackage{makecell}

\usepackage[round,authoryear]{natbib}
\usepackage[colorlinks=true,
            linkcolor=blue,
            citecolor=blue,
            urlcolor=blue]{hyperref}

\usepackage[switch]{lineno}

\usepackage{xcolor}
\usepackage{enumitem}
\usepackage{siunitx}

\title{\bfseries On the limiting geometry of unsteady breaking waves subject to co-flowing wind: spectrally-informed versus locally-measured steepness}

\author{
\begin{minipage}{0.9\textwidth}
\centering
Rui Cao$^{1,2,3}$, Enrique M. Padilla$^{4}$, Xu Chen$^{1,2}$ and Adrian H. Callaghan$^{3,*}$\\[0.5em]
\small $^{1}$State Key Laboratory of Physical Oceanography, Ocean University of China, Qingdao 266100, China\\
\small $^{2}$College of Oceanic and Atmospheric Sciences, Ocean University of China, Qingdao 266100, China\\
\small $^{3}$Department of Civil and Environmental Engineering, Imperial College London, London SW7 2AZ, UK\\
\small $^{4}$Barcelona School of Industrial Engineering, Universitat Politècnica de Catalunya (UPC), Barcelona 08034, Spain\\[0.5em]
\small $^{*}$Corresponding author: \href{mailto:a.callaghan@imperial.ac.uk}{a.callaghan@imperial.ac.uk}
\end{minipage}
}

\date{3 December 2026}

\begin{document}
\maketitle
\begin{abstract}
Wave steepness is a key geometric variable for describing the occurrence of breaking and its consequences like energy dissipation and air entrainment. Using three laboratory campaigns under varying spectral conditions and co-flowing wind forcing, we contrast two types of steepness commonly employed for unsteady breaking waves: spectrally-informed wave-group steepness (prognostic), obtained from fixed-point surface-elevation records, and locally-measured crest steepness (diagnostic), obtained from surface spatial profiles extracted using the SDBW-I image-processing method developed herein. For the former, the long-adopted $\mathcal{S}_n$ (linear sum of Fourier-component steepness) increases appreciably within approximately two dominant wavelengths upstream of breaking because of its strong sensitivity to evolving high-frequency content. When measured sufficiently far upstream, however, wave-group steepness remains approximately linearly related to the local zero-crossing steepness $\mathcal{S}_b$ under unforced conditions. Notwithstanding this, we argue that the crest-front steepness, $\mathcal{S}_{\mathrm{front}}(t_b)$, which delineates the slope of the front face at incipient breaking, is the most physically-meaningful metric among the steepness measures studied here. Specifically, within the conditions examined, it exhibits a consistent breaking-onset (lower-bound) threshold of $\mathcal{S}_{\mathrm{front}}(t_b)\approx0.2$, while its values above this threshold decrease with increasing wind speed (and decreasing wave age) as crests become less forward leaning. This may be attributed to the combined effects of wind-modified dispersion, enhanced high-frequency spectral content and aerodynamic sheltering, suggesting that wind--wave and non-linear wave--wave interactions act as competing mechanisms in triggering breaking through kinematic and energetic processes beyond what geometry can explain. Even so, $\mathcal{S}_{\mathrm{front}}(t_b)$ has strong potential as a controlling variable for studies of breaking energetics and crest-scale dynamics.
\end{abstract}


\section{Introduction}
\label{sec: intro}

\subsection{Broad motivation and background} \label{subsec: motivation}
Wave breaking is a discrete process occurring at the crests of sufficiently steep water waves, often in a seemingly random pattern in time and space. From a local perspective, breaking can be viewed as the exceedance of a (geometric, kinematic, or dynamic) threshold beyond which the crest can no longer maintain its form or grow further \citep[e.g.][]{Johannessen2001,Wu2002,Stansell2002,Banner2007,Toffoli2010,Knobler2022,Boettger2024,Hulin2025}. In a broad context, identifying this limiting state when breaking occurs has implications for understanding air-sea exchanges of momentum and mass (e.g. generation of bubbles and sea spray), wave field development (e.g. limiting wave heights and wave energy dissipation), and turbulence generation that enhances upper ocean mixing \citep{Melville1996,Duncan2001,Na2020,Deike2021,McAllister2024}. Among the possible descriptors of this limiting state, wave steepness remains one of the most widely used geometric measures, although its definition and interpretation are not unique for unsteady breaking waves.

In the ocean, breaking rarely occurs without wind. Co-flowing wind supplies momentum and energy to the wave field through surface stresses associated with form drag and skin friction, and can thereby modify wave growth, crest steepening, and the occurrence of breaking \citep{Miles1957,Hanson1999,Olfateh2017, Babanin2018, Yousefi2024, Scapin2025}. Around sufficiently steep crests, the overlying airflow may also separate, producing pressure asymmetry between the windward and leeward faces and altering the local forcing experienced by the crest \citep{Jeffreys1926,Touboul2006,Buckley2016,Lee2017,Lee2020,Kristoffersen2021, Tan2025,Buckley2025}. It is therefore reasonable to expect the limiting crest geometry at breaking to differ from that in unforced wave groups. However, because individual breaking events are difficult to isolate in the field, and because controlled wind-forced experiments remain relatively limited, how wind modifies the crest shape of unsteady breaking waves is still not well quantified. 

The broad aim of the present work is therefore to examine how the limiting crest geometry of unsteady breaking waves varies across different unforced and wind-forced conditions, using wave steepness as the primary geometric measure.

\subsection{Narrowing the problem} \label{subsec: review}

The definition of wave steepness, $\mathcal{S}$, usually involves a vertical and a horizontal length scale, but the particular choice of these scales varies across studies. Since most wind-driven oceanic breakers occur in unsteady wave groups, two approaches (types) have been developed to examine the breaking-wave steepness, as outlined below.

\begin{figure*}
    \centering
	\includegraphics[width=1\columnwidth]{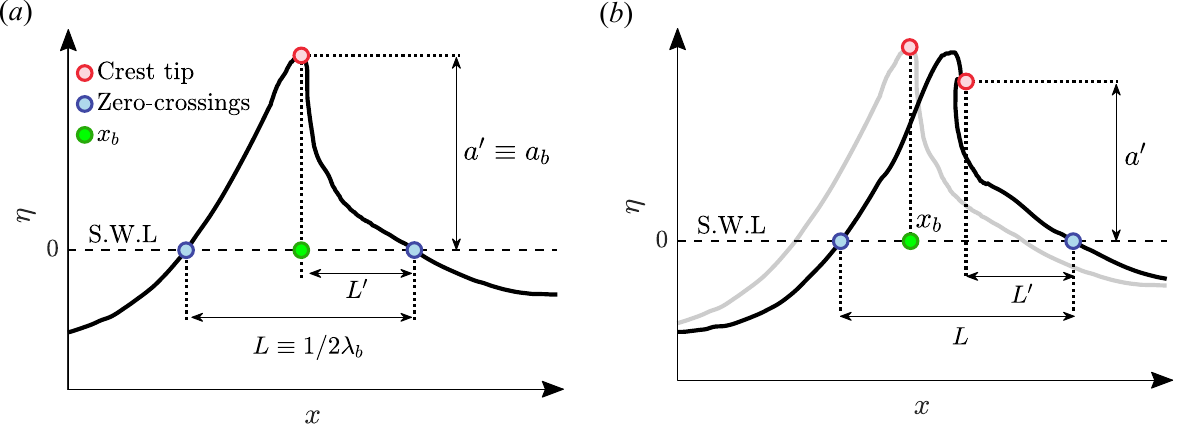}
	\caption{Illustrations of the parameters defining local crest geometry at (\textit{a}) incipient breaking and (\textit{b}) overturning. The wave propagates in the $+x$ direction. The profiles shown are from a real breaking event where its vertical dimension is exaggerated by a factor of 6.5 for clarity. In both panels, $x_b$ marks the crest location at the brink of breaking, $L$ the half-wavelength between zero crossings, and $a^{\prime}$ and $L^{\prime}$ the crest amplitude and crest-front distance, respectively, such that at incipient breaking $L=1/2 \, \lambda_b$ and $a^{\prime}=a_b$. The abbreviation, S.W.L, indicates the still water level.}
    \label{fig: LocalSteepnessDefinition}
\end{figure*}

The first approach provides a diagnostic measure of the limiting crest geometry at incipient breaking\footnotemark \footnotetext{This quantifies the limiting steepness attained by a crest at the instant of breaking, $t_b$, when a vertical front face is present.}. Following \citet{Derakhti2020} and \citet{Boettger2023}, we distinguish this moment from breaking inception, which marks the initiation of irreversible processes within the crest that lead to breaking, before any visible manifestation of breaking occurs. Measurements at incipient breaking therefore capture the limiting crest geometry in a wave that will break but before the crest overturns \citep{McAllister2023,McAllister2024}. A commonly employed local measure is the zero-crossing steepness,

\begin{equation}
\mathcal{S}_b=a_b \, k_b \equiv a_b\left(2 \pi /\lambda_b\right) \label{subeq:Sb},
\end{equation}
where $a_b$, $k_b$ and $\lambda_b$ are the local amplitude, wavenumber, and wavelength at incipient breaking, respectively (figure~\ref{fig: LocalSteepnessDefinition}\textit{a}). Reported values of $\mathcal{S}_b$ for unsteady breakers can be below the Stokes limiting steepness $\mathcal{S}\approx0.44$ for steady deep-water waves and can also vary with wave-group properties \citep{Tian2010,Banner2014crestspeed,Derakhti2016,Saket2017,Barthelemy2018}. Thus, although $\mathcal{S}_b$ is useful, it does not define a unique geometric breaking-onset threshold\footnotemark 
\footnotetext{This refers to the threshold magnitude of a steepness parameter required to trigger the least energetic breaking event and is often used as a prognostic criterion for breaking.}. Since breaking is often tied to the development of a forward-leaning crest \citep{Fedele2020}, other local measures concerning crest asymmetry and front-face slope may therefore be more directly connected to the limiting crest form \citep{Kjeldsen1979,She1994,Tian2008,Babanin2010,Perlin2013,DeVita2018,Chen2022,McAllister2023}. 

The second is what we refer to as wave-group steepness, computed from the wave group spectral composition measured upstream of the breaking location. This provides an $a \, priori$ (prognostic) measure of group non-linearity and is widely used in laboratory and numerical studies because it can be obtained from single-point surface-elevation records, without tracking the rapidly evolving crest profile \citep{Rapp1990,Drazen2008,Tian2010,Callaghan2013a,Deike2015,Deane2016,Pizzo2021,Sinnis2021,Xu2022,Cui2022,Cao2023,McAllister2024}. 

Not only insightful in describing the breaking-wave steepening geometry and hence the degree of non-linearity, locally-measured and spectrally-informed steepness have often been used interchangeably as controlling variables for breaking-wave dynamics, energetics, and air entrainment \citep[][and among many others]{Rapp1990,Erinin2023Profile}. In particular, wave-group steepness has been used to characterise the breaking strength of individual dispersively-focused laboratory breaking waves \citep{Drazen2008,Sinnis2021,Cao2023}, and has also been incorporated into spectral wave models to constrain the magnitude of the energy-dissipation source term \citep{Romero2019,Hogan2025}. These applications, however, implicitly assume that local crest steepness and spectrally-informed wave-group steepness are uniquely related.

Evidence for this assumption remains incomplete. One of the earliest studies in which a linear proportionality was reported came from \citet{Drazen2008}, but we note that the two types of steepness they compared were not entirely independent and therefore contained a degree of self-correlation (see their figure~15). On the other hand, a linear relationship between $\mathcal{S}_b$ and wave-group steepness ($\mathcal{S}_n$, defined later in equation~\eqref{eq:Sn}) was later reported in numerical and laboratory studies \citep{Tian2010,Deike2016}, but only for a limited set of dispersively-focused breaking waves with constant-steepness spectra. Whether the same correspondence holds for more realistic oceanic spectra, such as JONSWAP, and across varying underlying wave scales remains unresolved.

Furthermore, the action of direct wind forcing on wave group evolution leading to breaking may further complicate this correspondence. Indeed, laboratory experiments carried out in deep-water, wind-forced conditions have shown that sufficiently strong co-flowing wind can randomise the crest shape at incipient breaking \citep{Oh2005, Babanin2010,Saket2017}, with similar effects also reported numerically \citep{Chen2022, Boettger2024}. Otherwise, the wind influence on limiting wave steepness may not be measurable in general. For instance, the presence of positive vorticity, a possible consequence of wind-induced shear currents \citep{Xie2017}, has been shown to have little effect on limiting steepness \citep{Touboul2021}, although it may shift the breaking location downstream \citep{Chen2022}. These mixed findings motivate a systematic comparison of local and wave-group steepness under both unforced and wind-forced breaking conditions.

\subsection{Target and expectation} \label{subsec: target}

Our study makes several contributions toward understanding the limiting geometry of breaking surface waves. We first present novel experimental datasets acquired in a state-of-the-art wind-wave flume in which a broad range of wind and wave-group conditions were concerned (\S\ref{sec: setup}-\S\ref{sec: campaigns}). Second, we introduce in \S\ref{sec: profile extraction} an image-processing method for extracting overturning crest profiles and the spatial parameters required to quantify local steepness. These measurements are combined with in-situ wave-gauge data to examine the along-flume evolution of spectrally-informed wave-group steepness (\S\ref{subsec: Wave group steepness evolution}) and the temporal variation of locally-measured steepness (\S\ref{subsec: Local steepness evolution}), and to contrast the two types of steepness without (\S\ref{sec: comparison without wind}) and with (\S\ref{sec: comparison with wind}) superimposed, co-flowing winds.

By doing so, we address whether spectrally-informed and locally-measured steepness can be used interchangeably under unforced and wind-forced breaking conditions, and clarify the physical interpretation of each when used to characterise unsteady breaking waves. We will also show that crest-front steepness \eqref{subeq:Sf} provides a more physically meaningful diagnostic of the limiting crest form, with implications for how wave scale and wind forcing are represented in studies of unsteady breaking waves. Our paper concludes with a synthesis of these findings and their broader implications (\S\ref{sec: conclusion}).

\section{Experimental details and methods} \label{sec: exp}
\subsection{General flume configuration} \label{sec: setup}
We employed laboratory-scale analogues to study individual oceanic breaking waves under both unforced and wind-forced conditions. The findings in the present study comprise results from three experimental campaigns ({\bf SIREN}, {\bf BUBER}, and {\bf EURUS}), conducted between 2020 and 2021 in the unidirectional, wind--wave flume of the Hydrodynamics Laboratory within the Department of Civil and Environmental Engineering at Imperial College London. 

\begin{figure*}
    \centering
	\includegraphics[width=1\columnwidth]{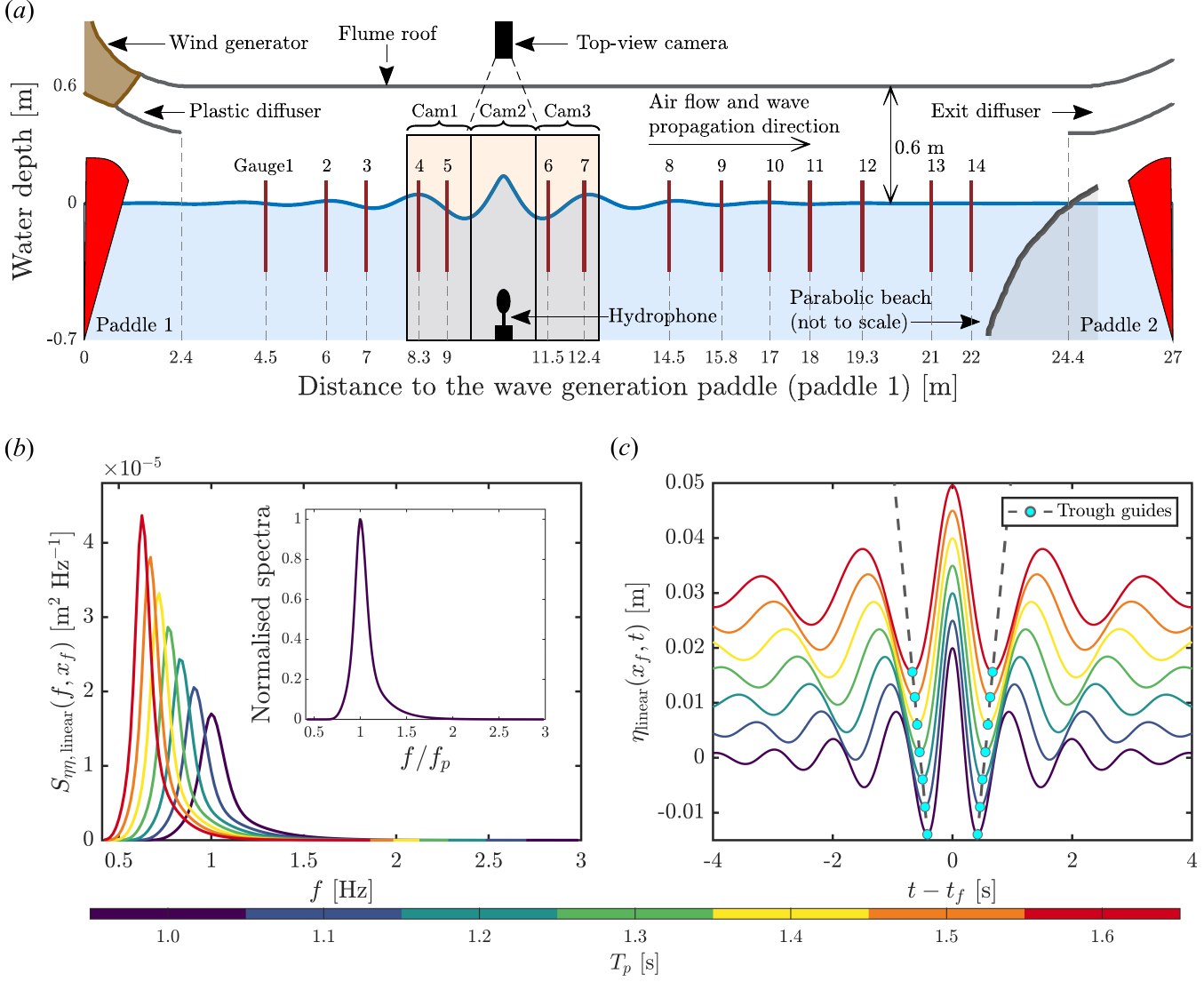}
	\caption{($a$) Schematic of the experimental set-up for \textbf{SIREN}. An earlier version of this panel was published in \citet{Cao2026Representative}. ($b$, $c$) Target (linear) variance density spectra, $S_{\eta\eta,\,\mathrm{linear}}(f_i,x_f)=a_i^2/(2\,\delta f)$, and corresponding linear surface elevations at the focal point for wave groups with equal $A=20$ mm, $\gamma=2$, and varying $T_p$. The inset in ($b$) shows the normalised spectra across different $T_p$ values, with frequency $f$ scaled by $f_p$ and variance density normalised by $S_{\eta\eta,\,\mathrm{linear}}(f_p,x_f)$. In ($c$), the surface elevations are vertically offset for clarity. Filled circles mark the adjacent troughs on either side of the target crest, and grey dashed lines provide visual guides through these troughs. The colourbar underneath panels ($b$) and ($c$) indicates the $T_p$ values used in \textbf{SIREN}.}
    \label{fig: SIREN setup}
\end{figure*}

The flume is 27 m long and 0.3 m wide, with thin transparent glass sidewalls for optical measurements. All experiments were conducted in freshwater at a fixed depth of $d=0.7$ m. As shown in figure~\ref{fig: SIREN setup}(\textit{a}), paddle 1 generated the intermediate-to-deep-water wave groups, while paddle 2 behaved as an active dissipater by which reflected wave energy was minimised. A parabolic beach with the space beneath it filled with plastic foam of 75\% porosity was installed next to paddle 2 to further reduce reflections.

Co-flowing wind was supplied at varying speed by a centrifugal wind generator mounted at the paddle-1 end of the flume. Air entered and exited through diffusers at both ends, with a vertical air space of 0.6 m between the still water level and the flume roof (although this height varied dynamically as waves propagated beneath the airflow).

\subsection{Wave-group generation} \label{sec: generation}
All three campaigns employed dispersively focused wave groups generated using linear theory, following approaches commonly used to produce unsteady, air-entraining breaking waves \citep{Rapp1990,Deane2002,Drazen2008,Sinnis2021,Cao2023, McAllister2024}. This technique allows a nominal focusing time ($t_f$) and focusing location ($x_f$) of a steep crest to be prescribed. The target surface elevation, $\eta(x,t)$, is given by linear superposition of $N$ freely propagating wave components, each of a specific frequency, $f_i$, and amplitude, $a_i$:

\begin{equation} \label{eq:eta}
\eta(x, t) =\sum_{i=1}^{N} a_{i} \cos \left[k_{i} \left(x-x_f\right)-2 \pi f_{i} \left(t-t_f\right)\right],
\end{equation}
where each frequency $f_i$ is linked to its wavenumber $k_i$ through the finite-depth dispersion relation, $(2\pi f_i)^2=k_i g \tanh{(k_id)}$, with $g$ the gravitational acceleration. The target crest amplitude at the focal point is therefore the linear amplitude sum of all spectral components, $\eta_{\text{linear}}(x=x_f,t=t_f)=A=\sum_{i=1}^{N} a_i$.

The component amplitudes are obtained from JONSWAP variance
density spectra following the \textit{NewWave} formulation \citep{Cao2023, Tromans1991}:

\begin{subequations}\label{eq: NewWave amplitudes}
\begin{align}
a_i &=
2S_{\mathcal{J}}(f_i)\,\delta f\,
\frac{A}{\displaystyle\sum_{j=1}^{N}2S_{\mathcal{J}}(f_j)\,\delta f}
=
A\frac{S_{\mathcal{J}}(f_i)}{\displaystyle\sum_{j=1}^{N}S_{\mathcal{J}}(f_j)},
\label{eq: NewWave scaling}
\\[6pt]
S_{\mathcal{J}}(f_i) &=
\frac{0.0081g^2}{(2\pi)^4 f_i^5}
\exp\!\left[-1.25\left(\frac{f_p}{f_i}\right)^4\right]
\begin{cases}
\displaystyle
\gamma^{\exp\!\left[-\frac{(f_i-f_p)^2}
{2(0.07f_p)^2}\right]}, & f_i\leq f_p, \\[6pt]
\displaystyle
\gamma^{\exp\!\left[-\frac{(f_i-f_p)^2}
{2(0.09f_p)^2}\right]}, & f_i>f_p,
\end{cases}
\label{eq: JONSWAP}
\end{align}
\end{subequations}
where $f_p$ is the spectral peak frequency and $\delta f$ is
the frequency spacing.

Thus, physical variables characterising the wave groups include the spectral peak period ($T_p=1/f_p$) and JONSWAP peak enhancement factor ($\gamma$) in equation~\eqref{eq: JONSWAP}, which were
prescribed independently to vary wave scale and underlying
spectral bandwidth, respectively. Specifically, increasing $T_p$ leads to an increased wavelength of the dominant spectral component, indicative of a larger wave scale \citep{Callaghan2013a, Cao2026Representative}; decreasing $\gamma$ broadens the spectrum where the wave group is more dispersive and experiences a reduced degree of non-linear dispersion and wave-wave interaction compared to a correspondingly more narrowband wave group where frequency components have relatively more time to interact \citep{Baldock1996,McAllister2023}. In addition, wind forcing was applied at varying speeds, and the wind condition is expressed using the equivalent 10-m wind speed, $\overline{U_{10}}$, and the corresponding wave age, $C_p/\overline{U_{10}}$ \citep{Yousefi2024}, where $C_p=2\pi f_p/k_p$ is the linear phase speed at the
spectral peak frequency.

For each $(T_p,\gamma,U_{10})$ combination, the component
amplitudes were scaled using equation~\eqref{eq: NewWave scaling} to generate sequences of wave groups with progressively increasing $A$, ranging from quasi-linear, non-breaking waves to highly non-linear incipient breakers and ultimately plunging events (see table~\ref{tab:exp_conditions}). In our experiments, only wave groups with a maximum of one breaking wave were analysed. When more than one breaking event occurred, $A$ was either further increased to reproduce a single breaking event, in which case the crest that broke could differ from that in the preceding cases, or the previous single-breaking case was retained. Although we note that the actual breaking location and time differed from the prescribed $(x_f,t_f)$ and varied with $A$ and wind speed, $x_f$ and $t_f$ in equation~\eqref{eq:eta} were tuned so that each breaking event occurred within the field of view of Cam2 (see figure~\ref{fig: SIREN setup}\textit{a}).

All wave groups used a repeat time of 64 s, giving a frequency resolution of $\delta f=1/64$ Hz. The underlying spectral range extended from 0.406 Hz ($26/64~\mathrm{Hz}$, fixed) to $3/T_p$ $(=3f_p)$, so that the number of components was $N=\left\lfloor(3/T_p-26/64)/\delta f\right\rfloor+1$, where $\lfloor\cdot\rfloor$ denotes the floor function that rounds its argument down to the nearest integer. For example, a wave group with $T_p=1.3$ s comprised $N=122$ individual spectral components.

For ease of reference, where needed, individual cases are labelled as x-G$\bullet$Tp$\bullet\bullet$A$\bullet\bullet\bullet$, in which the prefix denotes the campaign: {\bf SIREN} (S), {\bf BUBER} (N), and {\bf EURUS} (H, for the highest wind speed applied). As an example, S-G3Tp13A020 refers to a wave group from {\bf SIREN} with $\gamma = 3$, $T_p=1.3$ s, and $A=20$ mm.

\begin{table}
    \centering
\small
\setlength{\tabcolsep}{3.5pt}
\renewcommand{\arraystretch}{1.15}
    \resizebox{\textwidth}{!}{%
\begin{tabular}{ccccccccc}
\toprule
\makecell{Campaign\\name} 
& $T_p$ [s] 
& $\gamma$  
& \makecell{$A$ [$10^{-3}$ m]\\non-breaking} 
& \makecell{$A$ [$10^{-3}$ m]\\breaking} 
& \makecell{$\overline{U_{10}}$\\{}[m\,s$^{-1}$]} 
& \makecell{$C_p/$\\$\overline{U_{10}}$} 
& $N$ 
& $\tanh(k_p d)$ \\
\midrule 
\addlinespace
    \textbf{SIREN}    & 1.0 & 2   & 20-55  & 56-80  & n/a & n/a & 167 & 0.993 \\
                      & 1.1 & $-$ & 20-56  & 64-110 & $-$ & $-$ & 149 & 0.982 \\
                      & 1.2 & $-$ & 20-60  & 73-118 & $-$ & $-$ & 135 & 0.964 \\
                      & 1.3 & $-$ & 20-81  & 96-146 & $-$ & $-$ & 122 & 0.942 \\
                      & 1.4 & $-$ & 20-109 & 119-143  & $-$ & $-$ & 112 & 0.919 \\
                      & 1.5 & $-$ & 20-113 & 134-154  & $-$ & $-$ & 103 & 0.891 \\
                      & 1.6 & $-$ & 20-135 & 149-159  & $-$ & $-$ & 95 & 0.857 \\
                      
                      & 1.1 & 3   & 20-57  & 64-110  & $-$ & $-$ & 149 & 0.982 \\
                      & 1.3 & $-$ & 20-88  & 94-140  & $-$ & $-$ & 122 & 0.942 \\
                      & 1.5 & $-$ & 20-124 & 130-154 & $-$ & $-$ & 103 & 0.891 \\
\addlinespace
    \textbf{BUBER}    & 1.2 & 2 & 40-69  & 80-105 & 0   & No wind & 135 & 0.964 \\
                      & $-$ & 3 & 40-71  & 83-117 & $-$ & $-$     & $-$ & $-$ \\
                      & 1.3 & 2 & 40-75  & 90-130 & $-$ & $-$     & 122 & 0.942 \\
\addlinespace
    \textbf{EURUS}    & 1.2 & 2 & 40-69  & 80-105 & [3.0; 6.0] & [0.61; 0.30] & 135 & 0.964 \\
                      & $-$ & 3 & 40-71  & 83-117 & $-$ & $-$ & $-$ & $-$ \\
                      & 1.3 & 2 & 40-75  & 90-130 & [3.0; 6.0] & [0.64; 0.32] & 122 & 0.942 \\
    \end{tabular}
}
    \caption{Summary of input (linear target) experimental conditions for \textbf{SIREN}, \textbf{BUBER} and \textbf{EURUS} campaigns. 
    Columns indicate the spectral peak period ($T_p$), JONSWAP peak enhancement factor ($\gamma$), linear amplitude sum ($A$) for non-breaking and breaking cases, equivalent 10-m wind speed ($\overline{U_{10}}$), local wave age ($C_p/\overline{U_{10}}$), 
    number of spectral components contained in a wave group ($N$), and finite-depth metric ($\tanh(k_p d)$).}
    \label{tab:exp_conditions}
\end{table}

\subsection{Wave-group specifications and measurements} \label{sec: campaigns}
\subsubsection{SIREN}
The {\bf SIREN} campaign was conducted in 2020 to investigate how breaking depends on wave scale. 

The experimental arrangement is shown in figure~\ref{fig: SIREN setup}(\textit{a}). Along the flume centreline, 14 resistive wave gauges were deployed to record surface elevation at 128 Hz, of which five were placed upstream of the focal (breaking) zone and nine downstream. Three complementary metal--oxide--semiconductor (CMOS) cameras from the JAI GO series (Cam1--Cam3) imaged the crest evolution and the two-phase flow generated by breaking. Cam3 (the most downstream camera) was a 2.35-megapixel camera (1936 pixels $\times$ 1216 pixels, with active area 11.3 mm $\times$ 7.13 mm and cell size 5.86 $\mu$m $\times$ 5.86 $\mu$m) equipped with a Tamron 12 mm lens and Cam1 and Cam2 were 5.1-megapixel cameras (2464 pixels $\times$ 2056 pixels, with active area 8.50 mm $\times$ 7.09 mm and cell size 3.45 $\mu$m $\times$ 3.45 $\mu$m) equipped with Tamron 8 mm lenses. Together they covered approximately 4.5 m in the streamwise direction, from $x \simeq 8.0$ to 12.7 m. Illumination was provided by custom-built LED panels mounted opposite the cameras.

The cameras and light panels were synchronised using an Arduino-controlled trigger and operated at 52 Hz over a continuous 10 s recording window. This field of view and recording duration were sufficient to capture the full breaking process for the energetic groups considered here. The top-view camera and hydrophone shown in figure~\ref{fig: SIREN setup}($a$) are included for completeness but are not used in the present analysis (see \citet{Cao2026Representative} for further details).

The conditions examined are summarised in table~\ref{tab:exp_conditions}. Key parameters were varied over $T_p \in $ [$1.0, 1.6$] s, $\gamma\in\{2,3\}$, and $A$, which was progressively increased from 0.02 m until a controlled single breaking event could no longer be maintained within the field of view of Cam2. Figures~\ref{fig: SIREN setup}($b$, $c$) illustrate the target (linear) variance density spectra and corresponding linear surface elevations for a representative subset of cases with $A=20$ mm and $\gamma=2$. As $T_p$ increases the spectral peak shifts to lower frequency and the time-domain profile stretches horizontally (as illustrated by the increasing separation between the troughs on either side of the central crest in figure~\ref{fig: SIREN setup}\textit{c}), indicative of larger-scale wave groups. The inset in figure~\ref{fig: SIREN setup}($b$) shows that the normalised spectra have the same shape across different $T_p$ values, so that in a linear sense changes in the underlying wave scale are introduced without altering the prescribed spectral form.

\subsubsection{BUBER and EURUS}
Two further campaigns were conducted in 2021: \textbf{BUBER}, under unforced conditions, and \textbf{EURUS}, with direct co-flowing wind. They were designed to examine how wind forcing modifies the breaking process under otherwise comparable wave group conditions ($T_p\in\{1.2,1.3\}$ s, $\gamma\in\{2,3\}$; table~\ref{tab:exp_conditions}).

\begin{figure*}
    \centering
	\includegraphics[width=1\columnwidth]{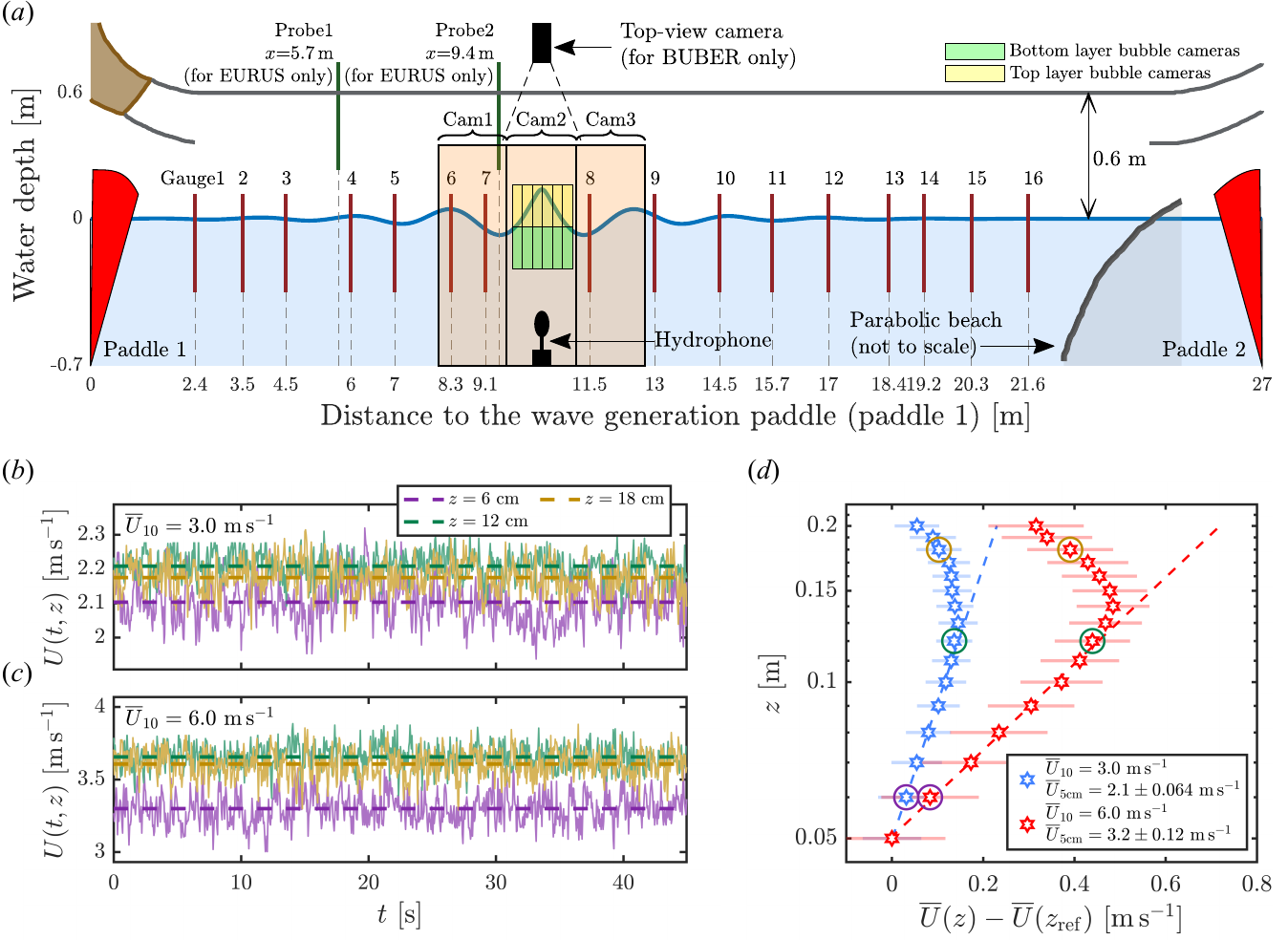}
	\caption{(\textit{a}) Schematic of the experimental set-up for \textbf{BUBER} and \textbf{EURUS}. As in figure~\ref{fig: SIREN setup}(\textit{a}), all instruments are shown, although the hydrophone, top-view camera, and bubble-camera data are not used here. The top-view camera was used only in \textbf{BUBER}, while the two wind probes, located 5.7 m and 9.4 m from paddle 1, were used only in \textbf{EURUS}. (\textit{b}, \textit{c}) Time histories of wind speed measured by probe 2 at $z=0.06$, $0.12$ and $0.18$ m above the still water level for extrapolated $\overline{U_{10}}=3.0$ and $6.0$ m$\,$s$^{-1}$, respectively. Solid lines show the measured velocities and dashed lines their corresponding time averages over 45 s, with colours indicating measurement height. (\textit{d}) Vertical profiles of mean wind speed, $\overline{U}(z)$, plotted relative to the measured reference speed at $z_{\mathrm{ref}}=0.05$ m, for the two \textbf{EURUS} wind conditions. Horizontal bars indicate $\pm$ one temporal standard deviation, and dashed lines show logarithmic fits. Reference speeds are given in the legend. Coloured circles mark the heights shown in (\textit{b}, \textit{c}), with matching colours.}
    \label{fig: BURUS setup}
\end{figure*}

The overall arrangement in \textbf{BUBER} was similar to \textbf{SIREN}, with several amendments (figure~\ref{fig: BURUS setup}\textit{a}). Sixteen wave gauges were installed along the flume, with seven upstream of the breaking zone and nine downstream. Cam1 and Cam3 were 2.35-megapixel cameras fitted with Tamron 12 mm lenses, while Cam2 was a 5.1-megapixel camera fitted with a Tamron 8 mm lens. Additional 5.1-megapixel cameras were employed to image bubble plumes (regions shaded by yellow and green colours), although those data are not analysed here. All cameras operated at 20 Hz with LED backlighting.

For \textbf{EURUS}, the flume was fully sealed to prevent air leakage along the flume between the air inflow and outflow locations. Two air-velocity transducers (TSI 8455-075-1), sampled at 20 Hz, were installed at fetches of 5.7 m and 9.4 m from paddle 1. Wind speed measurements from the downstream probe were used because this location was immediately upstream of the breaking zone and therefore more representative of the local airflow conditions. Before each focused wave group was generated, the wind was allowed to blow for 8 minutes to establish a statistically steady background wave field with small, high-frequency ripples (see also appendix~\ref{appA: wind-induced waves} for measurements on these purely wind-induced wave fields).

The characteristic wind speed is taken as the equivalent 10-m velocity ($\overline{U_{10}}$), extrapolated from logarithmic profiles measured under statistically steady, wind-only conditions:

\begin{eqnarray} \label{eq: U}
U(x,z)=\frac{u_{*}}{\kappa}\ln{\left[\frac{z}{z_0(x)}\right]},
\end{eqnarray}
where $\kappa=0.41$ is the non-dimensional von K\'arm\'an constant, $u_{*}$ is the friction velocity, and $z_0$ is the roughness length. The values of $u_{*}$ and $z_0$ were computed by fitting equation~\eqref{eq: U} to the measured wind profiles.

Figures~\ref{fig: BURUS setup}(\textit{b, c}) show wind speed time histories measured by probe 2 at three representative heights for the two \textbf{EURUS} wind conditions, $\overline{U_{10}}=3.0$ and $6.0$ m$\,$s$^{-1}$, respectively. The records show temporal fluctuations about their corresponding means (dashed lines), calculated over 45 s. The mean profiles are shown in figure~\ref{fig: BURUS setup}(\textit{d}), with horizontal bars indicating one temporal standard deviation on either side of the mean. Measurements were taken from $z=0.05$ to 0.2 m, avoiding positions too close to the free surface to protect the probe and reduce Wave Boundary Layer (WBL) effects \citep{Chalikov2011}. Because the wind speed decreases above approximately $z=0.13$ to $0.14$ m (consistent with boundary-layer development over the wavy water surface and along the flume roof and sidewalls; see also figure~3.11 of \citet{Gray2021}), only measurements at $z\leq0.13$ m were used for the logarithmic fits and for extrapolating $\overline{U_{10}}$.

Alongside underlying wave group conditions, the airflow conditions for \textbf{BUBER} and \textbf{EURUS} are summarised in table~\ref{tab:exp_conditions} using $\overline{U_{10}}$ and the wave age $C_p/\overline{U_{10}}$. For the wind-forced cases in \textbf{EURUS}, $\overline{U_{10}}=3.0$ and $6.0$ m$\,$s$^{-1}$ correspond to $C_p/\overline{U_{10}}\approx0.61$--$0.64$ and $0.30$--$0.32$, respectively, hereafter approximated as $0.6$ and $0.3$ for simplicity. Although mechanically generated wave groups may perturb the airflow, especially at larger amplitude and steepness, these values are used as representative indicators of the intended young and very young wind-wave regimes. 

Before presenting the results for breaking wave groups, we first examine in appendices~\ref{appA: wind-induced waves}--\ref{appA: wind-forced non-breaking waves} purely wind-induced waves and wind-forced non-breaking groups, through which the physical context for interpreting breaking-wave results is established (appendix~\ref{appA: summary}).

\subsection{Extraction of surface profile at incipient breaking and overturning} \label{sec: profile extraction}
Quantifying local wave steepness requires the air--water interface to be extracted accurately from side-view images. In our earlier work \citep{Cao2025}, we introduced the Continuous Maximum Gradient (CMG) method, in which the free surface is treated as a smooth streamline of connected pixels. This method has demonstrated good accuracy for non-breaking waves, with an RMSE of about 0.5 mm against independent wave-gauge measurements reported in \citet{Cao2025} (see also its applications in \citet{Almalki2025,Ehlers2025,Xu2025}), but becomes inadequate once the crest overturns. This is mainly because the interface now contains near-vertical segments and multi-valued profiles that break down the streamline hypothesis we made in CMG.

\begin{figure*}
    \centering
	\includegraphics[width=0.92\columnwidth]{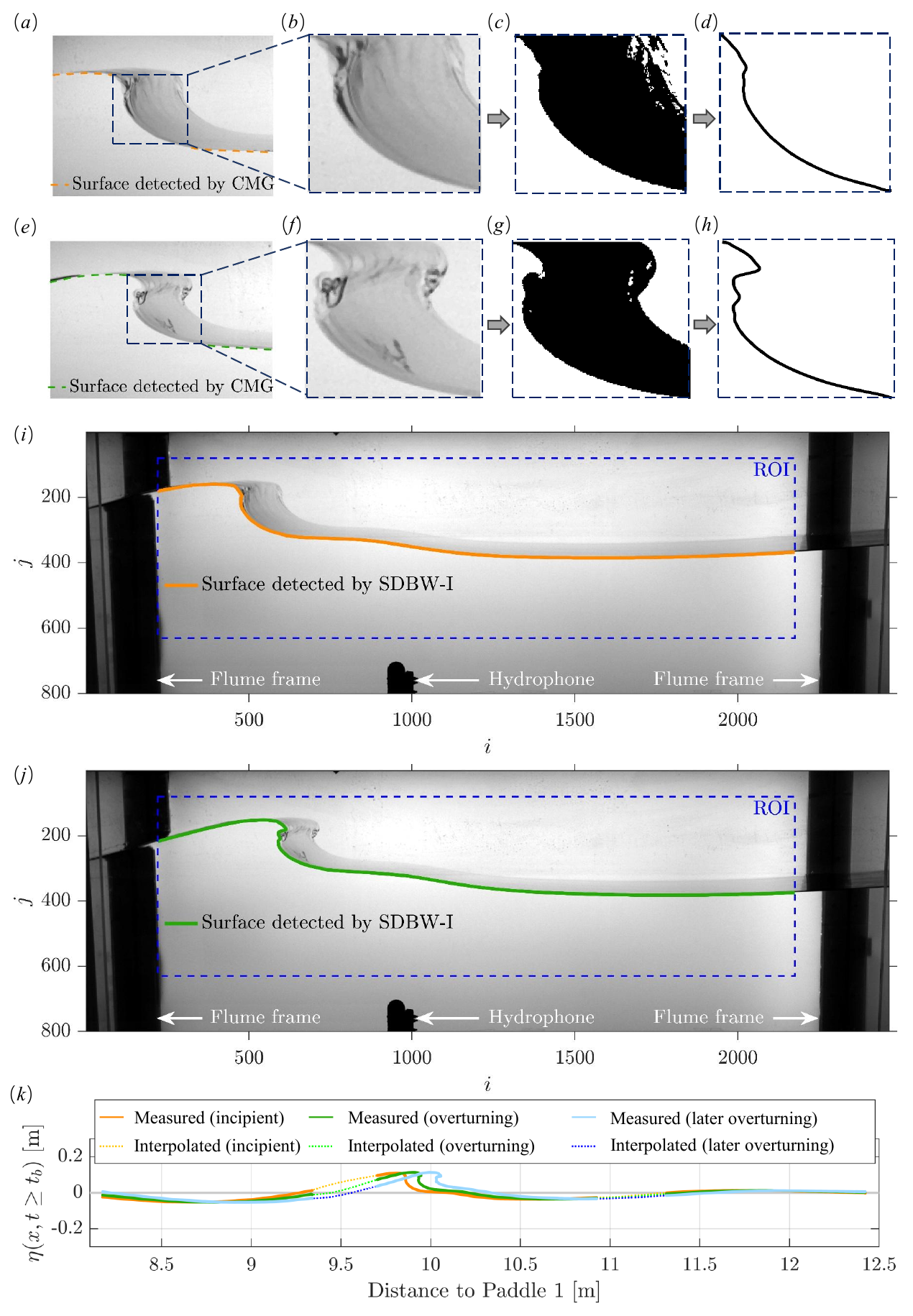}
	\caption{Mechanics and performance of SDBW-I. The first and second rows correspond to incipient breaking (vertical front face) and overturning, respectively: (\textit{a}, \textit{e}) snapshots of case S-G2Tp12A095 with partial CMG detection; (\textit{b}, \textit{f}) zoomed-in regions; (\textit{c}, \textit{g}) binary images from Otsu thresholding; and (\textit{d}, \textit{h}) extracted air–water interface along the inner face of the crest. (\textit{i}, \textit{j}) SDBW-I detection compared with Cam2 imagery, in pixel $i$-$j$ coordinates. (\textit{k}) Combined surfaces from three cameras in real-world coordinates with equal axes scales. Dotted lines denote the interpolation \citep[$S$-interp, ][]{Padilla2023} across gaps obscured by flume frames, as seen in figures~\ref{fig: SDBW - I demo}(\textit{i, j}).}    
    \label{fig: SDBW - I demo}
\end{figure*}

In resolving this limitation, we implement Phase I of the Surface Detection for Breaking Waves method (SDBW-I) developed in \cite{Cao2024}. The full SDBW framework distinguishes four optical phases of breaking; only Phase I is required here as it resolves the transition from incipient breaking, where a near-vertical front face is present, to early crest overturning. 

The procedure of SDBW-I is illustrated in figure~\ref{fig: SDBW - I demo}. After image calibration and pre-processing following \citet{Cao2025}, a rectangular Region of Interest (ROI) is defined around the steepening or overturning crest, while the neighbouring non-breaking portions of the free surface are still detected using CMG (figures~\ref{fig: SDBW - I demo}\textit{a}, \textit{e}). Under stable illumination, the interface at incipient breaking exhibits similar optical features across cases, allowing the same Otsu thresholding procedure to be applied to each ROI to obtain a binary image (figures~\ref{fig: SDBW - I demo}\textit{b}, \textit{f} $\rightarrow$ figures~\ref{fig: SDBW - I demo}\textit{c}, \textit{g}). The multi-valued interface is subsequently identified from the pixels forming the inner face of the crest (figures~\ref{fig: SDBW - I demo}\textit{c}, \textit{g} $\rightarrow$ figures~\ref{fig: SDBW - I demo}\textit{d}, \textit{h}). Since the waves propagate rightward, this interface is consistently taken as the left-hand boundary of the binary crest region. Combining this SDBW-I detection with the non-breaking portions extracted by CMG yields the complete surface profile within the camera field of view (figures~\ref{fig: SDBW - I demo}\textit{i}, \textit{j}).

The interface detected is then mapped from pixel coordinates $(i,j)$ to physical coordinates $(x,z)$ to obtain the surface profile. This transformation accounts for perspective effects arising from the non-perpendicular alignment of the camera optical axis relative to the $x$--$z$ plane of the wave flume, which introduces position-dependent scaling across the image. Lens distortion is removed through intrinsic calibration, while the remaining perspective-induced scaling variations are corrected using modified extrinsic parameters, following the procedures described in \citet{Cao2024}. This provides spatially consistent surface profiles across the measurement domain.

Figure~\ref{fig: SDBW - I demo}(\textit{k}) shows the transformed surface elevations from the three cameras at incipient breaking and during overturning. Gaps between adjacent camera views, caused by the flume metal struts, are interpolated using $S$-interp \citep{Padilla2023}. In our experiments, each breaking event was reconstructed independently three times, with the extracted interfaces carefully checked against the original images. The main geometric parameters (e.g. $a_b$, $L$ and $L^{\prime}$ defined in figure~\ref{fig: LocalSteepnessDefinition}) showed mean percentage variations of the order of $1\%$ or less across the three reconstructions. The uncertainties in the corresponding local wavenumber and steepness measures can be estimated accordingly. This gives confidence in the local steepness measurements used to examine their temporal evolution in \S\ref{subsec: Local steepness evolution} and their relationships with upstream wave-group steepness in \S\ref{sec: steepness comparison}.

\section{Evolution of wave-group steepness and local steepness} \label{sec: Steepness evolution}
\subsection{Wave-group steepness} \label{subsec: Wave group steepness evolution}
Spectrally-informed measures of wave-group steepness are derived from surface-elevation time histories, but their values can depend on where the measurements are made. As a focused wave group approaches breaking, rapid spectral evolution (especially in the high-frequency components) can modify the steepness inferred from the measured spectrum \citep{Derakhti2014,Cao2023}.

In our previous work \citep{Cao2023}, we compared several definitions of wave-group steepness between a far upstream location and a point immediately upstream of breaking where non-linearity was strong. That study, however, was limited in wave scale and did not include wind forcing. Here we focus on three metrics commonly used \citep{Cao2023}, briefly reintroduced below, and extend the analysis by examining their along-flume evolution across a broader range of wave scales and under direct co-flowing wind.

\subsubsection{Definitions and interpretation}
The first is the linear superposition of the steepness of all spectral components,

\begin{equation} \label{eq:Sn}
\mathcal{S}_n=\sum_{i=1}^{N} a_{i} \, k_{i}. 
\end{equation} 
Introduced by \citet{Melville1994}, $\mathcal{S}_n$ has been widely used as a prognostic measure of (linear) maximum steepness in deterministic breaking-wave studies \citep[e.g.][]{Drazen2008,Sinnis2021}.

The second metric scales the linear amplitude sum with the wavenumber at the spectral peak, 

\begin{equation} \label{eq:Sp}
\mathcal{S}_p=k_{p} \, \sum_{i=1}^{N} a_{i} \equiv k_{p} \,A, 
\end{equation}
where $k_p$ corresponds to the peak frequency $f_p$ through the linear dispersion relation. This definition is most appropriate for wave groups with a well-defined spectral peak \citep{Cao2023}.

The third metric incorporates a spectral energy-weighted frequency $f_s$:

\begin{equation} \label{eq:fs}
 f_s=\frac{\displaystyle \sum_{i=1}^{N} \left( f_i \, a_{i}^2 \right) \left( \delta f\right)_i}{\displaystyle \sum_{i=1}^{N} \left(a_{i}^2 \right) \left( \delta f\right)_i}, 
\end{equation}
with the corresponding steepness defined as:

\begin{equation} \label{eq:Ss}
\mathcal{S}_s=k_{s} \, \sum_{i=1}^{N} a_{i} \equiv k_{s} \,A, 
\end{equation}
where $k_s$ is the wavenumber associated with $f_s$. Previous studies have shown that $\mathcal{S}_s$ scales approximately linearly with the locally measured breaking steepness $\mathcal{S}_b$ (equation~\eqref{subeq:Sb}) \citep{Tian2010,Cui2022}. Compared with $\mathcal{S}_n$ and $\mathcal{S}_p$, it has also been found to provide stronger predictive skill for breaking onset and improved collapse when characterising breaking strength, owing to its explicit dependence on the spectral energy distribution \citep{Derakhti2016,Cao2023}.

\begin{figure*}
    \centering
	\includegraphics[width=1\columnwidth]{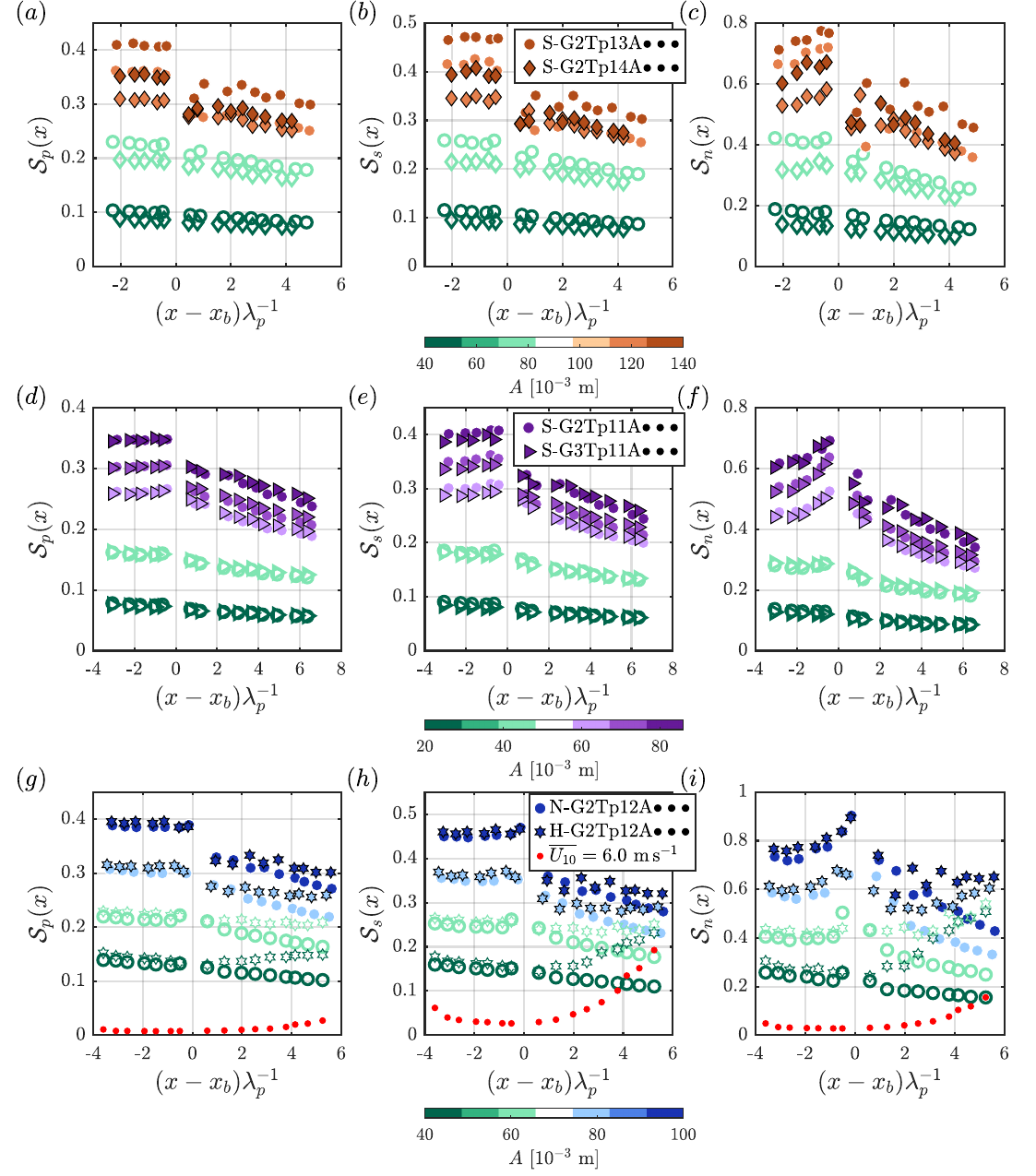}
	\caption{Spatial evolution of the three spectrally-informed wave-group steepness measures upstream and downstream of the focal/breaking locations. In each panel, hollow and filled symbols denote non-breaking and breaking wave groups, respectively, with colour indicating the linear amplitude sum $A$ (the dots $\bullet\bullet\bullet$ following $A$ in the legend labels denote the amplitude given by the colourbar). Panels (\textit{a})--(\textit{c}) show \textbf{SIREN} cases with $\gamma=2$ and varying wave scale ($T_p=1.3$ and $1.4$ s); panels (\textit{d})--(\textit{f}) show \textbf{SIREN} cases with fixed $T_p=1.1$ s and different spectral bandwidths; and panels (\textit{g})--(\textit{i}) show \textbf{BUBER} and \textbf{EURUS} cases with $\gamma=2$, $T_p=1.2$ s, and either no wind or co-flowing wind ($C_p/\overline{U_{10}}\approx0.3$). Red dots in panels (\textit{g})--(\textit{i}) denote steepness values of purely wind-generated waves at $\overline{U_{10}}=6.0$ m$\,$s$^{-1}$.}
    \label{fig: wave group evo}
\end{figure*}

\subsubsection{Effects of wave scale ($T_p$) and bandwidth ($\gamma$)}
Figure~\ref{fig: wave group evo} illustrates the spatial evolution of $\mathcal{S}_p$, $\mathcal{S}_s$, and $\mathcal{S}_n$ for a selection of wave groups with varying underlying wave scale, represented here by the spectral peak period $T_p$ (\textit{a}--\textit{c}), spectral bandwidth (\textit{d}--\textit{f}), and wind-forcing conditions (\textit{g}--\textit{i}). Both breaking and non-breaking cases are included. The streamwise coordinate is expressed relative to the focal/breaking location $x_b$ and normalised by the spectral peak wavelength $\lambda_p$, such that negative and positive values of $(x-x_b)\lambda_p^{-1}$ correspond respectively to upstream and downstream measurements.

A clear dependence on wave scale is observed in figures~\ref{fig: wave group evo}(\textit{a})--(\textit{c}). For otherwise identical conditions, wave groups with larger $T_p$ have smaller values of all three steepness metrics. This follows from the shift of the spectral peak to lower frequency as $T_p$ increases: the dominant wavelength becomes longer, and the same linear amplitude sum $A$ therefore corresponds to a less steep wave group. Thus, prescribing the same $A$ does not produce comparable wave-group steepness when the underlying wave scale changes. This $T_p$ dependence is apparent in both the breaking and non-breaking cases. 

By contrast, varying $\gamma$ within the limited range tested here produces no clear change in the wave-group steepness evolution (figures~\ref{fig: wave group evo}\textit{d}--\textit{f}): at comparable $A$, the values and spatial evolution of each metric are similar for the two bandwidths. Although our present comparison is limited to $\gamma=2$ and $3$, the absence of a clear bandwidth dependence here is consistent with \citet{Cao2023}, who examined a wider range of $\gamma$ and also found no pronounced bandwidth dependence.

Over the upstream measurement range ($(x-x_b)\lambda_p^{-1}<0$) without wind forcing (figures~\ref{fig: wave group evo}\textit{a}--\textit{f}), the three metrics remain approximately constant for non-breaking cases. Note that the wave gauges are fixed in space, so differences in $\lambda_p$ and $x_b$ between cases lead to different normalised measurement positions $(x-x_b)\lambda_p^{-1}$ and their corresponding ranges across the panels. For breaking cases, $\mathcal{S}_p$ still shows little spatial variation and $\mathcal{S}_s$ shows a marginal increase (which can reasonably be neglected; see figure~\ref{fig: wave group compare}), whereas
$\mathcal{S}_n$ increases appreciably as breaking is approached within approximately two dominant wavelengths, $-2\lesssim(x-x_b)\lambda_p^{-1}<0$. This pronounced increase in $\mathcal{S}_n$ is due to its heavier weighting of high-frequency components and their rapid evolution during non-linear focusing, as discussed in \citet{Cao2023}. Immediately downstream of $x_b$, all three metrics decrease abruptly, consistent with breaking-induced energy loss, and then continue to decay as the wave groups evolve further downstream and sidewall frictional dissipation takes effect.

\subsubsection{Effect of wind}
 
We next consider the effect of co-flowing wind. Figures~\ref{fig: wave group evo}(\textit{g})--(\textit{i}) compare $\mathcal{S}_p$, $\mathcal{S}_s$, and $\mathcal{S}_n$ between unforced wave groups from \textbf{BUBER} and wind-forced wave groups from \textbf{EURUS}. It is seen that wind increases all three metrics, but only slightly upstream of $x_b$. This is because, over the relatively short fetch upstream of breaking, wind input adds limited energy within the spectral range used to compute these steepness metrics (see also figures~\ref{fig: BackgroundWaves speed} and \ref{fig: BackgroundWaves fetch} in appendix~\ref{appA: wind-induced waves}). For reference, we show in the red dots the corresponding steepness values of purely wind-generated waves at $\overline{U_{10}}\approx6.0$ m$\,$s$^{-1}$.

Downstream of $x_b$, the wind effect becomes more pronounced, especially for wave groups with smaller $A$ and at longer fetches (similar to those seen in purely wind-induced wave fields as represented by red dots). As $A$ increases and the mechanically generated wave groups become more energetic, the relative contribution from wind weakens, suggesting that the added wind input is less important compared with the paddle-wave energy. This reduced influence is also consistent with the background-wave suppression mechanism of \citet{Touboul2006}, in which long paddle-generated waves interact with but inhibit the growth of shorter wind waves. These shorter components are then more readily dissipated by sidewalls and, in some cases, fall outside the frequency band used in the steepness calculation. We return later in \S\ref{ch: steepness implications} to discuss how different metrics of wave group steepness respond to the choice of upper frequency limit.

\subsubsection{Global comparison}
\begin{figure*}
    \centering
	\includegraphics[width=1\columnwidth]{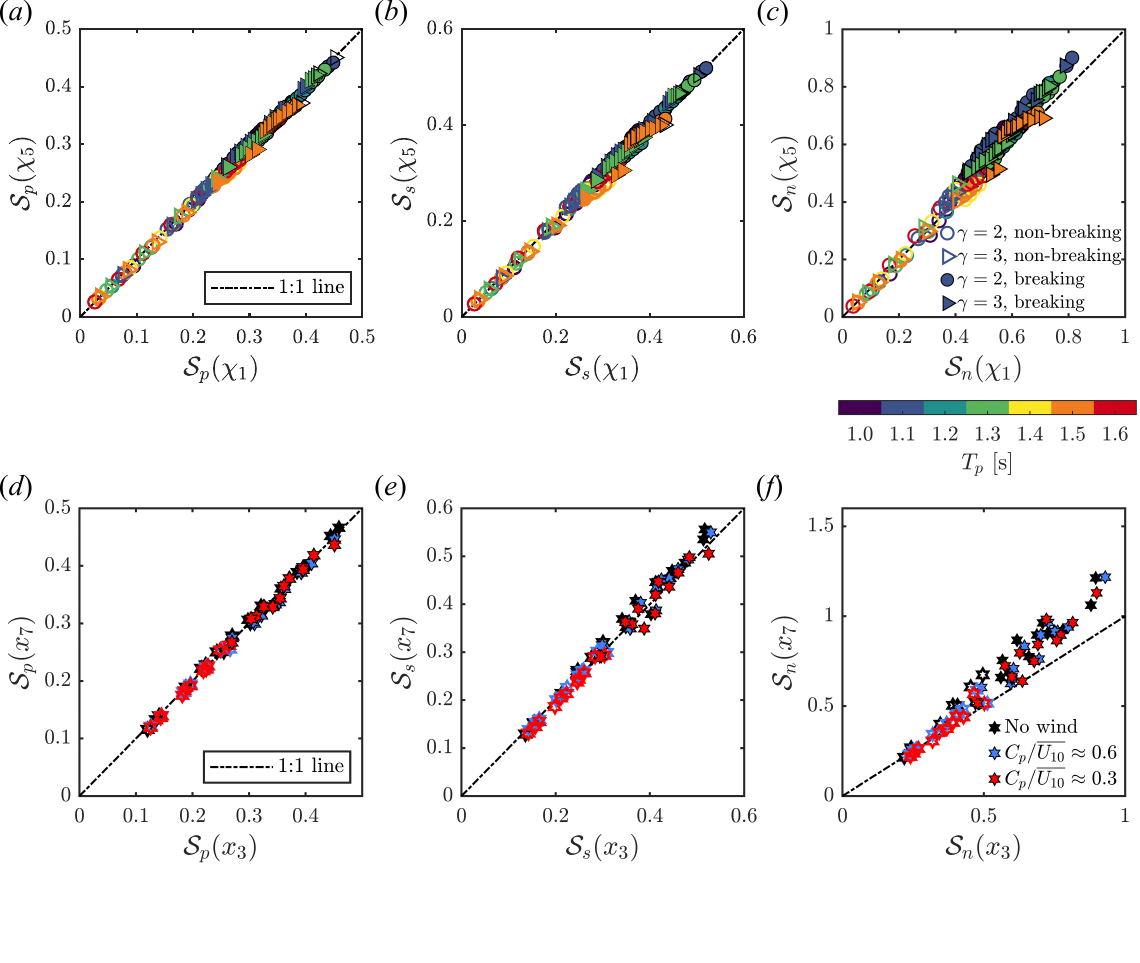}
	\caption{(\textit{a})--(\textit{c}) Comparisons of $\mathcal{S}_p$, $\mathcal{S}_s$, and $\mathcal{S}_n$ between the most upstream location ($\chi_1$) and a location just upstream of $x_b$ ($\chi_5$) for all non-breaking and breaking waves from \textbf{SIREN}. (\textit{d})--(\textit{f}) Equivalent comparisons for breaking cases from \textbf{BUBER} and \textbf{EURUS}, using $x_3$ and $x_7$, whose fetch distances are comparable to $\chi_1=4.5$ m and $\chi_5=9.0$ m in \textbf{SIREN} (see figures~\ref{fig: SIREN setup}\textit{a} and \ref{fig: BURUS setup}\textit{a}).}
    \label{fig: wave group compare}
\end{figure*}

We now compare wave-group steepness measured near the wavemaker with that measured just upstream of $x_b$ for the full dataset (figures~\ref{fig: wave group compare}\textit{a}--\textit{c} for \textbf{SIREN} and figures~\ref{fig: wave group compare}\textit{d}--\textit{f} for \textbf{BUBER} and \textbf{EURUS}). This tests whether the spatial trends identified above persist across different wave scales, bandwidths, and wind conditions. Note that we denote the \textbf{SIREN} gauges using $\chi$ to distinguish them from those in \textbf{BUBER} and \textbf{EURUS}; specifically, $\chi_1$ ($\simeq 4.5$ m) and $\chi_5$ ($\simeq 9.0$ m) have fetch distances comparable to $x_3$ and $x_7$, respectively (see also figures~\ref{fig: SIREN setup}\textit{a} and \ref{fig: BURUS setup}\textit{a}).

The results presented in figure \ref{fig: wave group compare} confirm that across the bulk of the data only $\mathcal{S}_n$ exhibits a measurable spatial dependence: steeper and thus more energetic wave groups experience a noticeable increase in $\mathcal{S}_n$ as they propagate towards $x_b$, whereas $\mathcal{S}_p$ and $\mathcal{S}_s$ remain comparatively insensitive to measurement location. This dependence is broadly independent of spectral bandwidth ($\gamma=2$ and $3$), wave scale ($T_p=1.0$--$1.6$ s), and wind forcing levels examined herein. Comparing figures~\ref{fig: wave group compare}(\textit{c}) and \ref{fig: wave group compare}(\textit{f}) also shows that $\mathcal{S}_n$ measured just upstream of $x_b$ is slightly higher for \textbf{BUBER} and \textbf{EURUS} than for \textbf{SIREN}. This is because the breaking location $x_b$ in \textbf{EURUS} and \textbf{BUBER} was shifted slightly upstream, demonstrating again the sensitivity of $\mathcal{S}_n$ to the non-linear evolution of propagating wave groups.

This sensitivity has a practical implication. If wave-group steepness is to be used as a prognostic indicator of local breaking steepness or breaking strength, it should not be evaluated too close to the breaking point. Under the present conditions, measurements taken more than approximately two dominant wavelengths upstream of $x_b$ are needed to avoid contamination by local non-linear growth (as shown in figure~\ref{fig: wave group evo}, right column).

\subsubsection{Implications for the use of different wave-group steepness metrics} \label{ch: steepness implications}

\begin{figure*}
    \centering
	\includegraphics[width=0.9\columnwidth]{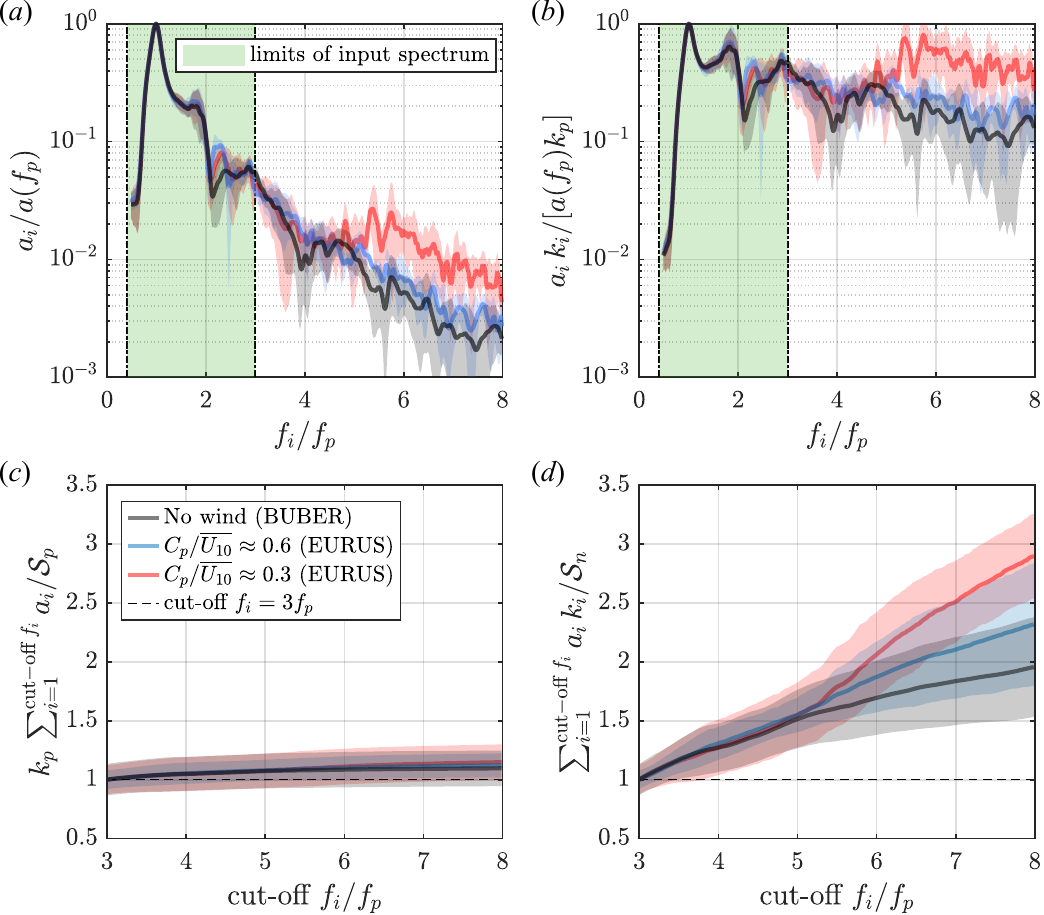}
	\caption{(\textit{a}, \textit{b}) Spectrally resolved distributions of $\mathcal{S}_p$ and $\mathcal{S}_n$ at $x_7=9.1$ m, just upstream of breaking. Values are normalised by their respective peaks, and shaded green regions mark the prescribed paddle-wave generation range. (\textit{c}, \textit{d}) Cumulative spectrally resolved wave-group steepness as the cut-off frequency is progressively extended, normalised by the reference values at $f_i/f_p=3$ (horizontal dashed lines). Results are ensemble averages of four G2Tp12 breaking events ($\gamma=2$, $T_p=1.2$ s) under unforced (\textbf{BUBER}) and wind-forced conditions ($C_p/\overline{U_{10}}\approx 0.6$ and $C_p/\overline{U_{10}}\approx 0.3$, \textbf{EURUS}). Shaded bands indicate $\pm$1 standard deviation from ensemble means.}
    \label{fig: spectral limits}
\end{figure*}

As noted earlier (and also described in appendices~\ref{appA: wind-forced non-breaking waves}--\ref{appA: summary}), upstream of $x_b$ the wind mainly increases the energy of the spectral components above the upper frequency limit used in the default steepness calculations ($f_N=3f_p$), where shorter wind-induced waves reside. This raises the question of whether the relative invariance in wave-group steepness between wind-unforced and wind-forced conditions is simply a consequence of restricting the integration band to $3f_p$.

To examine this, we consider $\mathcal{S}_p$ and $\mathcal{S}_n$ for four breaking events with $\gamma=2$ and $T_p=1.2$ s under different wind speeds ($\mathcal{S}_s$ is not shown because it behaves similarly to $\mathcal{S}_p$). Their spectrally resolved distributions at $x_7$ are shown in figures~\ref{fig: spectral limits}(\textit{a, b}). These distributions are normalised by their respective peak values, so for $\mathcal{S}_p$ it simplifies to $a_ik_p/[a(f_p) k_p]\rightarrow a_i/a(f_p)$. Two points stand out. First, within the prescribed paddle-wave band, wind has little effect on either metric, consistent with figures~\ref{fig: wave group evo}(\textit{g, i}). Second, a clear difference appears in the high-frequency tail. For $\mathcal{S}_p$ (figure~\ref{fig: spectral limits}\textit{a}), the wind-induced contribution remains one to two orders of magnitude smaller than that from the main band. For $\mathcal{S}_n$ (figure~\ref{fig: spectral limits}\textit{b}), however, the same high-frequency contribution can become comparable to, or even exceed, the main-band contribution because of the stronger weighting by higher wavenumber.

This contrast is further quantified in figures~\ref{fig: spectral limits}(\textit{c, d}), which show the cumulative spectrally-resolved steepness as we progressively extend the cut‑off frequency. The reference cut-off, $f_i=3f_p$, corresponds to the default range used to compute $\mathcal{S}_p$ and $\mathcal{S}_n$. Broadly speaking, the value of $\mathcal{S}_p$ is largely insensitive to this choice: extending the cut-off to $8f_p$ changes it by less than 10\%. By contrast, $\mathcal{S}_n$ increases rapidly, becoming about 50\% larger by $5f_p$. Beyond this point, wind effects become increasingly apparent; at $8f_p$, $\mathcal{S}_n$ under the strongest wind condition can be nearly twice its reference value at $3f_p$.

The implication here is worth emphasising. In the presence of wind forcing, unconstrained spectral integration (or summation) can imply artificially high wave-group steepness near breaking, even when this increase arises mainly from small but cumulative energy in the high-frequency tail. This represents an important limitation of spectrally-informed steepness metrics under wind, particularly for $\mathcal{S}_n$. We therefore turn next to locally-measured steepness (\S\ref{subsec: Local steepness evolution}) and examine how it compares with these spectral measures (\S\ref{sec: steepness comparison}).

\subsection{Local steepness}\label{subsec: Local steepness evolution}
As introduced in \S\ref{subsec: review}, local steepness differs from spectrally-informed (wave group) steepness in that it is a diagnostic parameter which, when measured at incipient breaking ($t=t_b$), naturally incorporates the combined effects of the pre-breaking processes that lead to breaking. Following equation~\eqref{subeq:Sb}, we first define a generic locally measured zero-crossing steepness from the geometric quantities shown in figure~\ref{fig: LocalSteepnessDefinition}: 

\begin{equation}
\mathcal{S}_{\mathrm{zc}}(t)= \frac{\pi a^{\prime}(t)}{L(t)} \label{eq:Szc}.
\end{equation}
At incipient breaking ($t=t_b$), $\mathcal{S}_{\mathrm{zc}}(t_b)\equiv \mathcal{S}_b$. Additionally, we define a more localised steepness for the crest front face:  

\begin{equation}
\mathcal{S}_{\text{front}}(t)= \frac{a^{\prime}(t)}{L^{\prime}(t)} \label{subeq:Sf},
\end{equation}
 which characterises the forward-leaning front face commonly observed as breaking is approached, when the front-face slope becomes much larger than that of the rear face (see \citet{Perlin2013} and the example in figure~\ref{fig: SDBW - I demo}).

\begin{figure*}
    \centering
	\includegraphics[scale=0.8]{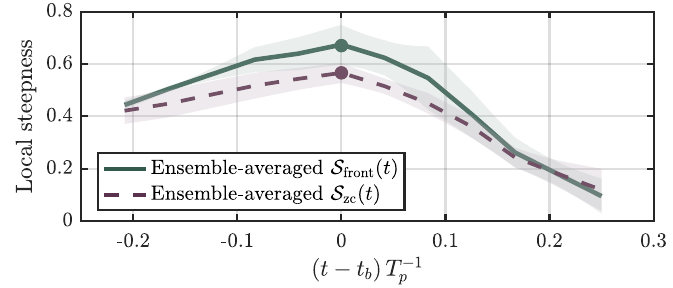}
	\caption{Time-evolving histories of the locally-measured zero-crossing steepness, $\mathcal{S}_{\text{zc}}(t)$, and crest-front steepness, $\mathcal{S}_{\text{front}}(t)$, before and after incipient breaking at $t_b$. Results are ensemble averages over five \textbf{BUBER} breaking waves with similar $\mathcal{S}_n$. Shaded regions indicate $\pm$1 standard deviation, and solid points at $(t-t_b)/T_p=0$ mark values of local steepness at incipient breaking.}
    \label{fig: local wave evo}
\end{figure*}

These local steepness metrics are quantified from the SDBW-I surface reconstructions described in \S\ref{sec: profile extraction}. Figure~\ref{fig: local wave evo} shows the temporal evolution of $\mathcal{S}_{\mathrm{zc}}$ and $\mathcal{S}_{\mathrm{front}}$ for an ensemble of five \textbf{BUBER} breaking waves with similar wave-group steepness $\mathcal{S}_n$.

Both local steepness metrics show a broadly similar trend: they  increase as the crest steepens towards breaking and decay rapidly after $t_b$. Their maxima occur at incipient breaking, consistent with local steepness acting as a diagnostic measure of crest evolution. After breaking, the crest loses its coherent form and no longer steepens. These results provide a useful reference for \S\ref{sec: steepness comparison}, where we compare local and wave-group steepness under different wave and wind conditions.

\section{Relationships between locally-measured and spectrally-informed steepness} \label{sec: steepness comparison}
\subsection{Comparison across wave scales in the absence of direct wind stress} \label{sec: comparison without wind}

\begin{figure*}
    \centering
	\includegraphics[width=1\columnwidth]{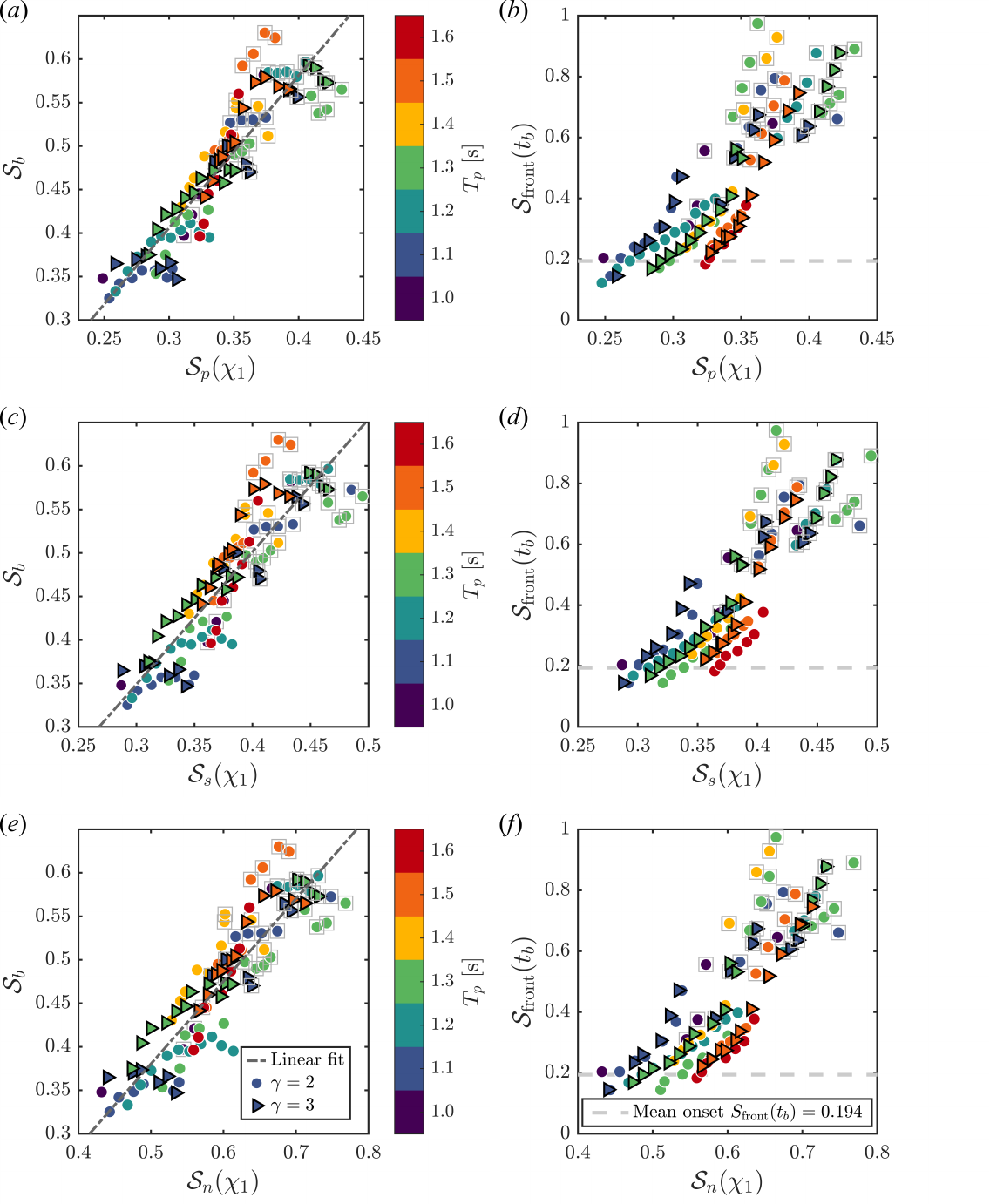}
	\caption{Comparisons between local crest steepness at incipient breaking ($\mathcal{S}_b$, left column, and $\mathcal{S}_\text{front}(t_b)$, right column) and upstream wave-group steepness ($\mathcal{S}_p(\chi_1)$, $\mathcal{S}_s(\chi_1)$ and $\mathcal{S}_n(\chi_1)$) for \textbf{SIREN} breaking cases across different wave scales and spectral bandwidths. Light grey square frames mark cases for which a further increase in $A$ restored a single breaking event but changed which wave within the group broke, as described in \S\ref{sec: generation}. Dash-dotted lines in the left panels show linear least-squares fits. Grey dashed lines in the right panels indicate the mean of the minimum measured values of $\mathcal{S}_{\mathrm{front}}(t_b)$ for the unforced conditions tested, i.e. the breaking-onset threshold.}
    \label{fig: steepness compare SIREN}
\end{figure*}

\subsubsection{$\mathcal{S}_b$ versus different measures of wave-group steepness}
In the left panels of figure~\ref{fig: steepness compare SIREN}, the locally-measured zero-crossing steepness at incipient breaking, $\mathcal{S}_b$, is compared with the three upstream wave-group steepness metrics, $\mathcal{S}_p(\chi_1)$, $\mathcal{S}_s(\chi_1)$, and $\mathcal{S}_n(\chi_1)$, for all \textbf{SIREN} breaking cases. Despite variations in $T_p$ and $\gamma$, $\mathcal{S}_b$ remains strongly correlated with all three metrics. Following \citet{Drazen2008} and \citet{Tian2010}, in which approximately linear relationships were reported, we quantify the corresponding trends in our data using simple empirical linear least-squares fits:

\begin{equation} \label{eq: Sb - Sp}
\mathcal{S}_{b}= (1.77\pm0.15)\mathcal{S}_p(\chi_1)-(0.13\pm0.05), 
\qquad R^2 = 0.83,
\end{equation}
\begin{equation} \label{eq: Sb - Ss}
\mathcal{S}_{b}= (1.52\pm0.15)\mathcal{S}_s(\chi_1)-(0.11\pm0.06), 
\qquad R^2 = 0.79,
\end{equation}

and
\begin{equation} \label{eq: Sb - Sn}
\mathcal{S}_{b}= (0.92\pm0.09)\mathcal{S}_n(\chi_1)-(0.08\pm0.05), 
\qquad R^2 = 0.81.
\end{equation}
Note that attempts were also made to relax the exponent of wave group steepness from its fixed value of 1, but this increased $R^2$ by only $0.003$--$0.011$, depending on the wave group steepness metric used. The near-linear trends obtained here broadly support previous reports of $\mathcal{S}_b$--$\mathcal{S}_n$ scaling \citep{Drazen2008,Deike2016} and $\mathcal{S}_b$--$\mathcal{S}_s$ scaling \citep{Tian2010,Cui2022}, although the exact fitted gradients and intercepts differ between studies, as do some of the steepness definitions, such as that of $\mathcal{S}_b$ in \citet{Drazen2008}.

Whilst these coefficients do not yet have a clear physical interpretation and may depend on the conditions and methods used in each study, the approximately linear trends reported in the literature and observed in the present data indicate that wave-group steepness measured sufficiently far upstream of breaking provides a useful way to characterise the local zero-crossing steepness at incipient breaking in the absence of wind forcing. This remains the case even for $\mathcal{S}_n$, despite its stronger sensitivity to high-frequency spectral content compared with $\mathcal{S}_p$ and $\mathcal{S}_s$. 

\subsubsection{$\mathcal{S}_{\text{front}}(t_b)$ versus different measures of wave-group steepness}
When the crest-front steepness at incipient breaking, $\mathcal{S}_{\mathrm{front}}(t_b)$, is used in place of $\mathcal{S}_b$, the influence of wave scale becomes evident (figure~\ref{fig: steepness compare SIREN}, right panels). Larger-scale wave groups with larger $T_p$ need to become more non-linear, as characterised here by higher spectrally-informed wave-group steepness, to attain the same level of $\mathcal{S}_{\mathrm{front}}(t_b)$ as lower-$T_p$ wave groups. We also note that at larger wave-group steepness, the data become more scattered (and the $T_p$-dependence is much less clear), partly associated with a change in which crest within the wave group breaks (cases enclosed by light grey square frames; see also explanations in \S\ref{sec: generation}). As a result, the spectrally-informed steepness measures do not collapse the limiting front-face geometry in the same way as they do for $\mathcal{S}_b$.

\begin{figure*}
    \centering
	\includegraphics[width=0.95\columnwidth]{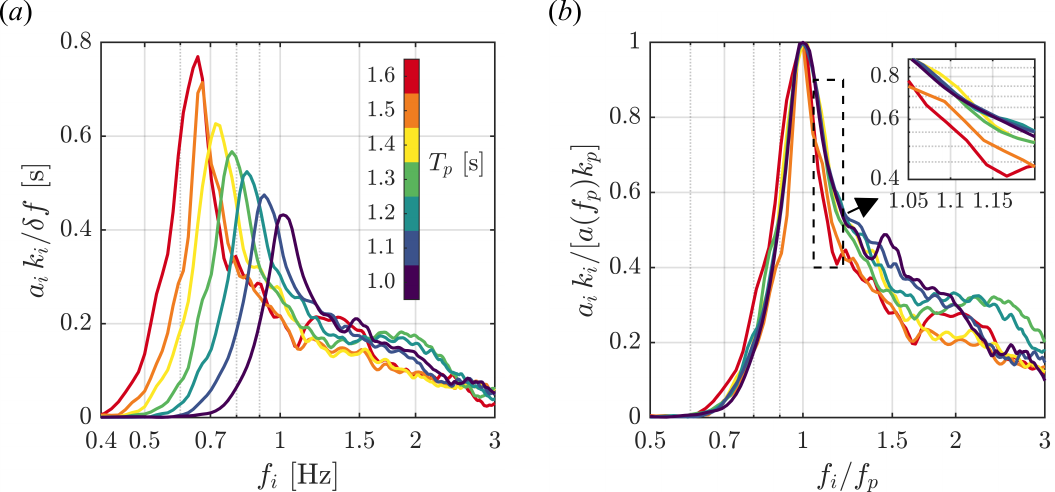}
	\caption{Measured spectrally-resolved steepness distributions of Fourier-decomposed components for the \textbf{SIREN} $\gamma=2$ (non-breaking) wave groups with varying $T_p$ but comparable wave-group steepness $\mathcal{S}_n(\chi_1)\in [0.30,\,0.35]$. Panel $(a)$ shows the absolute steepness density, $a_i k_i/\delta f$, as a function of frequency, and panel $(b)$ shows the corresponding distributions normalised by the spectral peak, $a_i k_i/[a(f_p)k_p]$, as a function of the normalised frequency $f_i/f_p$.}
    \label{fig: measured steepness distribution}
\end{figure*}

The above scale dependence may be explained by differences in the spectral steepness distributions among wave groups with different $T_p$. To demonstrate this we show in figure~\ref{fig: measured steepness distribution} selected non-breaking \textbf{SIREN} cases with comparable measured wave-group steepness $\mathcal{S}_n(\chi_1)\in [0.30,\,0.35]$ but different $T_p$, thereby elucidating the wave scale effect on the spectral steepness distribution. As $T_p$ decreases, the dominant steepness contribution shifts towards higher frequencies (figure~\ref{fig: measured steepness distribution}\textit{a}). This matters because high-frequency components are particularly relevant to crest instability and hence to the occurrence of breaking \citep{Cao2023}. Indeed, although the variance-density spectra are self-similar when normalised by their peaks (figure~\ref{fig: SIREN setup}\textit{b}, inset), the corresponding steepness-density spectra are not: smaller-$T_p$ groups have relatively larger high-frequency tails (figure~\ref{fig: measured steepness distribution}\textit{b}), allowing them to reach a given limiting $\mathcal{S}_{\mathrm{front}}(t_b)$ with a lower degree of wave group non-linearity.

Beyond wave scale, spectral bandwidth may also influence the limiting crest geometry, because it controls the duration of linear and non-linear interactions among the underlying components \citep{McAllister2023}. In the present comparisons, $\mathcal{S}_p$ provides the clearest test of such a bandwidth effect as it is independent of bandwidth by definition at fixed $A$ and $T_p$. No clear bandwidth dependence is apparent in the relation between $\mathcal{S}_{\mathrm{front}}(t_b)$ and $\mathcal{S}_p(\chi_1)$ (figure~\ref{fig: steepness compare SIREN}\textit{b}), but the limited variation between $\gamma=2$ and $3$ may be insufficient to resolve such an effect. A broader range of $\gamma$ would therefore be needed to assess this effect more fully.

The other robust result from figure~\ref{fig: steepness compare SIREN} (right panels) is that $\mathcal{S}_{\mathrm{front}}(t_b)$ exhibits a clear lower-bound threshold of approximately 0.2 across the unforced cases considered here. This threshold is not observed for $\mathcal{S}_b$ or for any of the spectrally-informed wave-group steepness metrics. This highlights the importance of accounting for the front--rear asymmetry of the crest when defining a breaking-onset criterion, as $\mathcal{S}_{\mathrm{front}}(t_b)$ does by construction. 

Taken together, our results suggest that, among the five steepness metrics studied here, $\mathcal{S}_{\mathrm{front}}(t_b)$ is better suited to describing the limiting local crest shape across
different underlying wave scales ($T_p$) at incipient
breaking in the absence of wind forcing and to defining a breaking-onset criterion (i.e. the lower limit for breaking) through its relatively invariant lower-bound value. Of course, the applicability of this lower bound beyond the present unidirectional, dispersively focused wave groups, with $\gamma=2$ and $3$, remains to be established.

\subsection{Comparison in the presence of direct wind stress} \label{sec: comparison with wind}
\subsubsection{Effect of wind on the steepness relationships}
We now ask whether the relationships between locally-measured and spectrally-informed steepness we saw under unforced conditions (figure~\ref{fig: steepness compare SIREN} in \S\ref{sec: comparison without wind}) also apply to breaking waves in the presence of co-flowing wind. 

\begin{figure*}
    \centering
	\includegraphics[width=1\columnwidth]{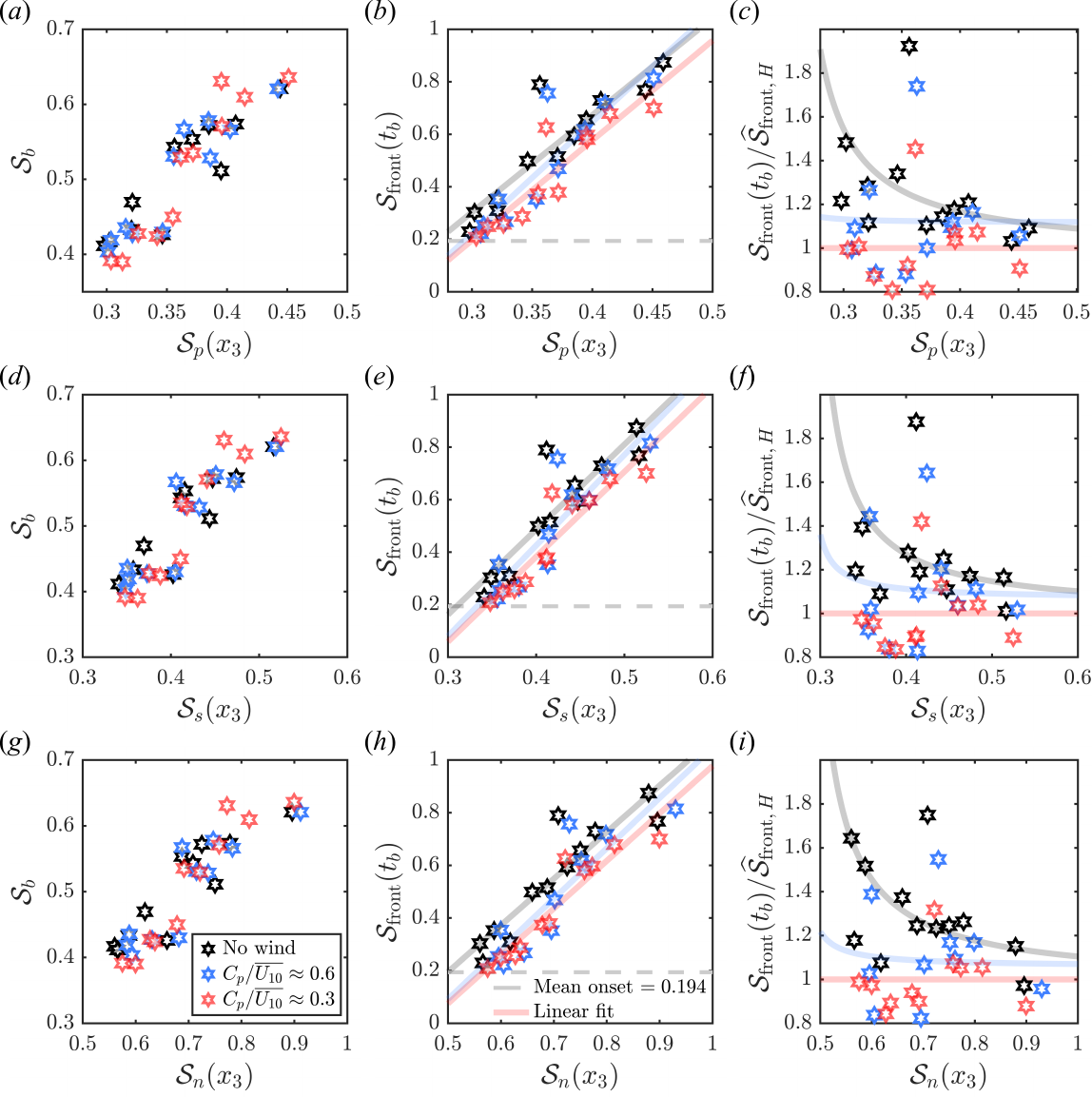}
	\caption{Comparisons of locally-measured crest steepness ($\mathcal{S}_b$ and $\mathcal{S}_{\mathrm{front}}(t_b)$) with spectrally-informed wave-group steepness ($\mathcal{S}_p(x_3)$, $\mathcal{S}_s(x_3)$, and $\mathcal{S}_n(x_3)$) for breaking cases from \textbf{BUBER} and \textbf{EURUS} under unforced conditions and co-flowing wind at $C_p/\overline{U_{10}}\approx0.6$ and $0.3$. Panels (\textit{a}, \textit{d}, \textit{g}) show $\mathcal{S}_b$, and panels (\textit{b}, \textit{e}, \textit{h}) show $\mathcal{S}_{\mathrm{front}}(t_b)$. Grey dashed lines mark the average breaking-onset threshold (i.e. the lower-bound value) identified from the unforced cases in figure~\ref{fig: steepness compare SIREN}. Solid lines in panels (\textit{b, e, h}) show independent linear least-squares fits for the three wind conditions, with colours matching the corresponding symbols. Panels (\textit{c, f, i}) show $\mathcal{S}_{\mathrm{front}}(t_b)$ normalised by the strongest-wind fit, $\widehat{\mathcal{S}}_{\mathrm{front},\,H}$, evaluated at the corresponding wave-group steepness. Solid curves show the fitted relations normalised by the same reference, so that red lines correspond to unity. Fit coefficients, their 95\% confidence intervals, and $R^2$ values are given in table~\ref{tab:fit_coefficients} of appendix~\ref{appB}.}
    \label{fig: steepness compare BURUS}
\end{figure*}

The left and middle columns from figure~\ref{fig: steepness compare BURUS} present the same comparison as figure~\ref{fig: steepness compare SIREN}, but now for \textbf{BUBER} and \textbf{EURUS}. It should be noted that in assessing the wind effect, $\mathcal{S}_p(x_3)$, $\mathcal{S}_s(x_3)$, and $\mathcal{S}_n(x_3)$ can be treated as upstream independent variables because wind input only weakly modifies these metrics at $x_3$ under the present conditions (cf. figure~\ref{fig: wave group evo}). Of course, this approximation would be less appropriate if wind input substantially contaminated frequencies close to the upper limit of the integration band.

The results in figures~\ref{fig: steepness compare BURUS}(\textit{a}, \textit{d}, and \textit{g}) reveal that $\mathcal{S}_b$ remains approximately linearly related to the three wave-group steepness metrics across the wind speeds considered. This suggests that, under the co-flowing wind conditions examined here, wind has only a weak influence on the local zero-crossing steepness at incipient breaking, a finding broadly consistent with the limited wind effect on maximum crest height reported by \citet{Touboul2006}.

A clear wind influence is seen for $\mathcal{S}_{\mathrm{front}}(t_b)$, however, as illustrated in figures~\ref{fig: steepness compare BURUS}(\textit{b}, \textit{e}, \textit{h}). Although the average threshold of $\mathcal{S}_{\mathrm{front}}(t_b)\approx0.2$ identified from the unforced cases still provides a useful breaking-onset reference (grey dashed lines), for a given value of wave-group steepness, wind-forced breaking appears to occur at smaller crest-front steepness. To make this reduction clearer, we independently fit the three wind conditions in figures~\ref{fig: steepness compare BURUS}(\textit{b}, \textit{e}, \textit{h}) using linear relations between $\mathcal{S}_{\mathrm{front}}(t_b)$ and each wave-group steepness metric. These fits are shown as solid lines with colours matching the corresponding data, and their fit statistics are provided in table~\ref{tab:fit_coefficients} of appendix~\ref{appB}. The expected values given by the strongest-wind fits ($C_p/\overline{U_{10}}\approx0.3$), $\widehat{\mathcal{S}}_{\mathrm{front},\,H}$, are then used to normalise all data and fitted relations in figures~\ref{fig: steepness compare BURUS}(\textit{c}, \textit{f}, \textit{i}). In this way, $\mathcal{S}_{\mathrm{front}}(t_b)/\widehat{\mathcal{S}}_{\mathrm{front},\,H}>1$ indicates crest-front steepness larger than expected from the strongest-wind relation. Notwithstanding the scatter in individual observations, the normalised data and fitted relations support a general reduction in $\mathcal{S}_{\mathrm{front}}(t_b)$ in the presence of wind, with a larger reduction generally observed at the higher wind speed, corresponding to the smaller wave age (i.e. younger breaking waves).

\begin{figure*}
    \centering
	\includegraphics[width=0.9\columnwidth]{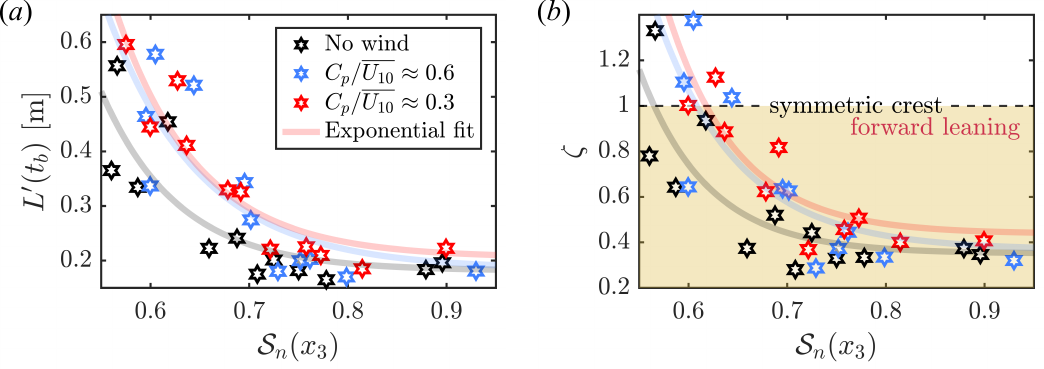}
	\caption{(\textit{a}) Crest-front distance $L^{\prime}(t_b)$ and (\textit{b}) crest asymmetry parameter $\zeta$ (defined in equation~\eqref{eq: lean}), plotted against $\mathcal{S}_n(x_3)$ for \textbf{BUBER} and \textbf{EURUS} breaking waves under varying co-flowing wind conditions. Solid curves show exponential fits for the three wind conditions, with colours matching the corresponding symbols. Fit coefficients, their 95\% confidence intervals and $R^2$ values are given in table~\ref{tab:fit_coefficients} of appendix~\ref{appB}. In (\textit{b}), the dashed line indicates $\zeta=1$ (corresponding to a symmetric crest), and the region shaded indicates the forward-leaning regime.}
    \label{fig: leaning}
\end{figure*}

We find in figure~\ref{fig: leaning}(\textit{a}) that relative to unforced cases, the wind-induced reduction in $\mathcal{S}_{\mathrm{front}}(t_b)$ is associated with an increase in the crest-front distance $L^{\prime}(t_b)$ at incipient breaking (see also figure~\ref{fig: LocalSteepnessDefinition}\textit{a} and equation~\eqref{subeq:Sf}). This motivates us to examine the relative balance between the front (negative slope) and rear (positive slope) portions of the crest using the non-dimensional asymmetry parameter

\begin{equation}
\zeta = \frac{L^{\prime}(t_b)}{L(t_b)-L^{\prime}(t_b)},
\label{eq: lean}
\end{equation}
defined following \citet{Kjeldsen1979}. Note that although this parameter was originally defined in the time domain, we apply it here in space. As shown in figure~\ref{fig: leaning}(\textit{b}), wind-forced cases generally have larger $\zeta$ within the forward-leaning regime ($\zeta < 1$) compared to unforced cases, implying that the crest leans less forward and is more symmetric at incipient breaking. 

To describe the observed trends in both $L^{\prime}(t_b)$ and $\zeta$, we fit each variable and wind condition independently using the empirical relation

\begin{equation}
\mathcal{V} = \mathcal{E}_3 + \mathcal{E}_1
e^{-\mathcal{E}_2[\mathcal{S}_n(x_3)-0.55]},
\qquad
\mathcal{V}\in\left\{L^{\prime}(t_b),\,\zeta\right\}.
\label{eq: crest geometry exponential fit}
\end{equation}
Here, $\mathcal{E}_3$ is the asymptotic value of the empirical curve, $\mathcal{E}_1$ is the departure from this value at $\mathcal{S}_n(x_3)=0.55$, and $\mathcal{E}_2$ controls the rate of decay with increasing $\mathcal{S}_n(x_3)$. Since the measurements show a degree of convergence at larger $\mathcal{S}_n(x_3)$, we fix $\mathcal{E}_3$ at the mean of the three observations with the largest $\mathcal{S}_n(x_3)$ for each fit, whilst $\mathcal{E}_1$ and $\mathcal{E}_2$ are estimated by least squares. The fitted curves are presented in figures~\ref{fig: leaning}(\textit{a}, \textit{b}), with fit results summarised in table~\ref{tab:fit_coefficients} of appendix~\ref{appB}.

The data and fitted curves shown in figure~\ref{fig: leaning}(\textit{b}) indicate that the wind-induced change in crest asymmetry (towards a less forward-leaning crest) appears strongest at lower $\mathcal{S}_n(x_3)$ and diminishes as $\mathcal{S}_n(x_3)$ increases, consistent with the convergence noted above. We interpret this to imply that a given level of wind forcing has a weaker relative effect on crest asymmetry for highly non-linear breaking wave groups compared to breaking wave groups with lower non-linearity. Here, higher wind speeds also correspond to smaller wave ages, so these trends should be interpreted within the relatively young wind-wave conditions examined. Further experiments covering a broader range of wave ages, including larger variations in wave speed at a fixed wind speed, would help assess how these trends extend to more mature wave fields.

\subsubsection{Discussion on the wind effect}
The wind-induced reduction in crest asymmetry and $\mathcal{S}_{\mathrm{front}}(t_b)$ is a striking result. In deterministic focused wave groups, forward leaning is usually understood as a kinematic outcome of dispersive focusing: longer waves catch up with shorter waves, the crest narrows and slows, and kinetic energy is converted into potential energy as the crest steepens \citep{Fedele2020}. For strongly non-linear groups, breaking tends to occur during this slowdown phase, when the crest is most asymmetric. Our present results therefore suggest that this pathway to breaking is influenced by direct wind-forcing. 

In addition, although the wave groups considered here were generated by dispersive focusing (i.e. freely propagating components are expected to coalesce at a prescribed phase point), the actual breaking process is also affected by non-linear wave--wave interactions that occur during the relatively short wave group propagation duration before breaking occurs \citep{Gibson2007}. It is within this context that we consider several coupled mechanisms through which wind may affect the limiting crest geometry. Below, we discuss how the phase changes and enhanced high-frequency spectral energy shown in our measurements (appendix~\ref{appA}), together with possible aerodynamic sheltering, may contribute to the reduced crest-front steepness:

\begin{enumerate}
\renewcommand{\labelenumi}{  (\roman{enumi}) }

\item As demonstrated in appendices~\ref{appA: wind-forced non-breaking waves} and \ref{appA: summary}, wind-induced drift currents impose disproportionate modulations on wave group dispersion between shorter and longer waves. Estimates based on representative exponential drift-current profiles suggest that shorter components are more strongly advected than longer components. This frequency-dependent change may modify the relative phase evolution and prolong the near-focusing state (i.e. the period over which the surface remains highly steepened but not necessarily in perfect phase alignment), potentially allowing instabilities to develop before the crest reaches the same degree of forward leaning as in the unforced case. \\ 

\item Wind also adds energy to the high-frequency part of the spectrum, as shown by the measurements in \S\ref{ch: steepness implications} and appendix~\ref{appA: wind-forced non-breaking waves}. Since breaking has long been associated with spectral saturation and the steepness of the high-frequency tail \citep{Phillips1985,Cao2023}, this additional (free) short-wave content may help promote breaking, although its contribution to the reduced limiting crest-front steepness is not directly quantified here. From appendix~\ref{appA: wind-forced non-breaking waves} it is seen that in weakly non-linear groups, the wind-induced components tend to superimpose linearly onto the mechanically generated spectrum (figure~\ref{fig: Wind Focused Spectra}). For highly non-linear wave groups their interaction with the underlying group becomes more complicated, including possible suppression or redistribution of energy to even shorter waves \citep{Touboul2006, Gray2021}. \\

\item Albeit not directly measured here, aerodynamic sheltering (most commonly referred to as Jeffreys' sheltering \citep{Jeffreys1926}) cannot be ruled out as an additional mechanism by which wind may advance breaking. As discussed in \S\ref{subsec: motivation} for sufficiently steep crests, airflow separation can generate pressure asymmetry between the windward and leeward faces \citep{Iafrati2019,Lee2020,Feddersen2023,Maleewong2024,Tan2025}. Such pressure perturbations can enhance particle velocities in the vicinity of the slowing wave crest and help satisfy a kinematic breaking condition earlier than in the unforced case \citep{Babanin2010,Saket2017,Zou2017,Chen2022,Zhang2024Transfer}. Support for this mechanism is provided by the numerical results of \citet{Boettger2024}, who compared breaking waves of similar initial steepness with and without wind forcing. They found that wind led to earlier breaking inception and an approximately 35\% increase in crest-tip kinetic energy relative to the unforced case before overturning. Similar observations were also reported by \citet{Xie2017}. \\ 

\item Finally, the larger $\mathcal{S}_{\mathrm{front}}(t_b)$ observed in unforced cases suggests that non-linear wave--wave interactions play a stronger role when wind is absent. In such cases, spectral energy around the largest crest can be redistributed into bound-wave components, increasing crest elevation, shortening the zero-crossing length scale, and producing a more asymmetric waveform \citep{Johannessen2003,Gibson2007,Gray2021}. Under wind forcing, by contrast, the dispersion modification, high-frequency input, and crest--airflow coupling mentioned above can lead to a relatively earlier occurrence of breaking, reducing the extent to which the crest must rely on non-linear amplification (and linear crest superposition) before breaking. In this sense, wind--wave interactions and non-linear wave--wave interactions may act as competing processes whose relative contributions influence the limiting steepness at incipient breaking. \\
\end{enumerate}

Whilst some of the mechanisms outlined above remain speculative and are not explicitly quantified here, the reduced limiting crest-front steepness under wind forcing can be reasonably interpreted as the combined effects of (i) wind-modified dispersion, (ii) enhanced high-frequency spectral content, and (iii) aerodynamic sheltering. Together, these effects suggest that breaking initiation is not set by geometry alone, whether measured spectrally or locally, but also requires energetic and kinematic considerations. Such considerations underpin the kinetic-energy growth criterion of \citet{Boettger2023} and the normalised energy-flux criterion of \citet{Barthelemy2018} and \citet{Derakhti2020}, which identifies imminent breaking (prior to incipient breaking) through the ratio of local energy flux to the product of energy density and crest speed, reducing at the free surface to the fluid-to-crest velocity ratio. The geometric information provided by $\mathcal{S}_{\mathrm{front}}(t_b)$ complements these approaches and may help relate the processes leading to breaking to the limiting crest shape.

\section{Summary of key insights and discussions} \label{sec: conclusion}
 The present study examined the limiting geometry of unsteady breaking waves using three laboratory campaigns (\textbf{SIREN}, \textbf{BUBER} and \textbf{EURUS}) conducted in a unidirectional wind--wave flume. The experiments covered varying wave scales, a limited range of spectral bandwidths, and co-flowing wind forcing. Combining in situ wave-gauge measurements with surface reconstructions from a novel image-processing technique introduced in this study (namely SDBW-I), we have compared and contrasted spectrally-informed (wave-group) steepness, used as an upstream prognostic measure, with locally-measured steepness, used as a diagnostic measure of the crest geometry at incipient breaking.

Our first result concerns the spatial evolution of wave-group steepness. The three metrics considered, $\mathcal{S}_p$, $\mathcal{S}_s$ and $\mathcal{S}_n$, remain approximately invariant when measured sufficiently far upstream of breaking. Closer to the breaking location, within approximately two dominant wavelengths upstream of $x_b$, $\mathcal{S}_n$ systematically increases, demonstrating its stronger sensitivity to high-frequency spectral content during non-linear focusing. Whilst this makes $\mathcal{S}_n$ useful for describing local wave-group (non-linear) evolution, caution is advised when using it as a predictive measure too close to the breaking location itself.

Our second result is that, in the absence of wind forcing, the local zero-crossing steepness $\mathcal{S}_b$ scales quasi-linearly with upstream wave-group steepness, in line with earlier findings reported in the literature \citep[e.g.][]{Deike2016, Tian2010}. Thus, when measured sufficiently far upstream, wave-group steepness may provide a useful proxy for estimating $\mathcal{S}_b$ at incipient breaking. This correspondence does not, however, extend in the same way to the crest-front steepness $\mathcal{S}_{\mathrm{front}}(t_b)$. Instead, $\mathcal{S}_{\mathrm{front}}(t_b)$ reveals a clear wave-scale dependence: larger-scale (i.e. higher $T_p$) wave groups require greater spectrally-informed steepness to reach the same limiting front-face geometry. This behaviour is consistent with their reduced high-frequency steepness density that can be important in helping trigger a breaking event.

The third result concerns the interpretation of steepness under wind forcing. Depending on the spectral cut-off, wave-group steepness $\mathcal{S}_n$ can be strongly affected by high-frequency components generated explicitly by wind-forcing. The local steepness definition avoids this added complexity because it is measured directly from the crest geometry at incipient breaking, rather than inferred from a spectrally integrated quantity. For a given upstream wave-group steepness, wind has limited effect on $\mathcal{S}_b$, but it systematically reduces $\mathcal{S}_{\mathrm{front}}(t_b)$ under the young wind-wave conditions examined here. This reduction is primarily associated with larger crest-front distances ($L^{\prime}$), as opposed to a smaller incipient wave amplitude, and weaker forward leaning at incipient breaking (quantified by a crest asymmetry parameter, $\zeta$). We interpret this as the combined effects of (i) wind-modified dispersion, (ii) enhanced high-frequency spectral content, and possibly (iii) aerodynamic sheltering, although their individual contributions are not quantified here.

Taken together, these considerations suggest that the presence of co-flowing wind, in a sense, reduces the extent to which wave breaking relies on non-linear energy transfer and focusing. More broadly these findings highlight that the initiation of breaking cannot be fully described by geometric considerations alone; energetic and kinematic processes must also be recognised. 

Finally, we conclude that among the five steepness metrics examined here, $\mathcal{S}_{\mathrm{front}}(t_b)$ captures differences in the limiting crest geometry associated with wave scale and wind forcing, suggesting that it may serve as a useful local parameter for characterising the subsequent energetic and dynamic processes in breaking waves. In addition, it exhibits an approximately consistent lower bound of $\mathcal{S}_{\mathrm{front}}(t_b)\approx0.2$ within the experimental conditions considered, which provides a geometric reference for breaking onset (i.e. the threshold for the least energetic breaking event). Whether this lower bound also applies to broader spectral bandwidths, other wind conditions and directionally spread wave fields is worth further investigation.

\section*{Acknowledgements}
The authors gratefully acknowledge the experimental assistance given by the technicians, Mr. David Ruyter and Mr. Gary Austin, of the Hydrodynamics Laboratory at the Department of Civil and Environmental Engineering, Imperial College London.

\section*{Funding}
The experimental campaigns reported herein, \textbf{SIREN}, \textbf{BUBER}, and \textbf{EURUS}, were funded by a Natural Environment Research Council (NERC) grant (Grant No. \href{https://gtr.ukri.org/projects?ref=NE\%2FT000309\%2F1}{NE/T000309/1}) awarded to A.H.C. in 2019 and conducted during R.C.'s doctoral studies at Imperial College London, during which he was supported by the Skempton PhD Scholarship. The work presented in this paper is supported by the National Natural Science Foundation of China (Grant No. 42606015), the China Postdoctoral Science Foundation (Grant No. 2025M770858), the Postdoctoral Innovation Program of Shandong Province (Grant No. SDCX-ZG-202601014), and the Qingdao Postdoctoral Science Foundation (Grant No. QDBSH20250202010), all awarded to R.C.

\section*{Author contributions}
R.C.: Project design, Conceptualisation, Data collection, Methodology, Data analysis, Writing - Original draft, Writing - Review \& Editing, Funding Acquisition; 
E.M.P.: Project design, Data collection, Writing - Review \& Editing; 
X.C.: Writing - Review \& Editing, Funding Acquisition, Supervision; 
A.H.C.: Project design, Conceptualisation, Writing - Review \& Editing, Supervision, Funding Acquisition, Final approval.

\section*{Declaration of interests}
 The authors report no conflict of interest.

\section*{Author ORCID} 
Rui Cao, https://orcid.org/0000-0002-2360-9412;  \\ 
Enrique M. Padilla, https://orcid.org/0000-0003-0532-8809; \\ 
Xu Chen, https://orcid.org/0000-0001-8526-9612;\\ 
Adrian H. Callaghan, https://orcid.org/0000-0001-7531-5072.

\appendix

\section{Wind-forced non-breaking wave fields}\label{appA} 
While the wind conditions in {\bf EURUS} were measured locally right above the breaking region (\S\ref{sec: campaigns}), the wind profile evolves along the fetch, leading to varying degrees of interactions with underlying wave fields. To place the breaking-wave results in context, it is therefore useful to first examine how co-flowing wind modifies non-breaking waves. The evolution of purely wind-induced wave fields and wind-forced, non-breaking focused wave groups has been examined in previous two-dimensional laboratory studies \citep{Hara1991,Touboul2006,Grare2013,Gray2021,Shemer2020,Shemer2021, Kumar2024}. In this appendix, we report our own measurements of both cases and interpret them using mechanisms established in those studies, providing the physical context for the wind-forced breaking results discussed in the main text.

\subsection{Purely wind-induced wave fields} \label{appA: wind-induced waves}
\begin{figure*}
    \centering
	\includegraphics[width=1\columnwidth]{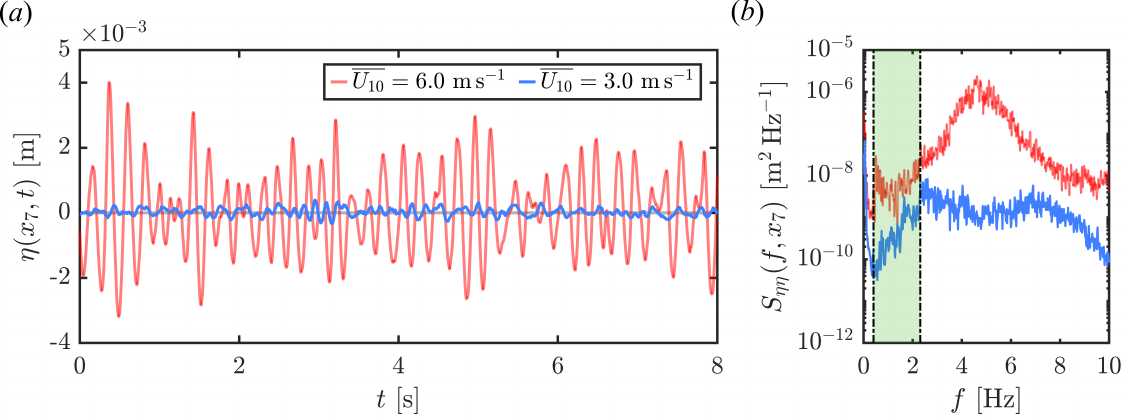}
	\caption{(\textit{a}) Surface elevation time-series of wind-induced wave fields at the two wind speeds used in {\bf EURUS}, upon reaching a statistically-stationary state. Measurements were taken at gauge 7 ($x=x_7$), the gauge closest to the focal/breaking location in \textbf{EURUS} (see also figure~\ref{fig: BURUS setup}\textit{a}). (\textit{b}) Corresponding FFT variance density spectra calculated from 256 s surface elevation records. The region enclosed by the vertical dashed lines represents the frequency range of the target spectra used to generate paddle waves.}
    \label{fig: BackgroundWaves speed}
\end{figure*}
Figure~\ref{fig: BackgroundWaves speed}(\textit{a}) shows examples of surface elevation time-series for wind-induced wave fields at statistically-steady state under $\overline{U_{10}}=6.0$ and $3.0$ m$\,$s$^{-1}$, measured near the breaking location in \textbf{EURUS}. The results demonstrate that background waves are generated by the overlying wind, but their development does not scale linearly with $\overline{U_{10}}$: at $\overline{U_{10}}=6.0$ m$\,$s$^{-1}$ a clear wave field develops, whereas at half this speed the waves are much weaker.

Figure~\ref{fig: BackgroundWaves speed}(\textit{b}) presents the corresponding FFT variance-density spectra, computed from 256 s surface-elevation records taken after more than 200 s of wind ramp-up and before paddle waves were generated. At this location, the stronger wind case ($\overline{U_{10}}=6.0$ m$\,$s$^{-1}$) shows increased spectral energy across the full frequency range, with a distinct peak near $f\approx5$ Hz, beyond the designed paddle-wave spectrum. No comparable peak is observed at the lower wind speed.

\begin{figure*}
    \centering
	\includegraphics[width=0.8\columnwidth]{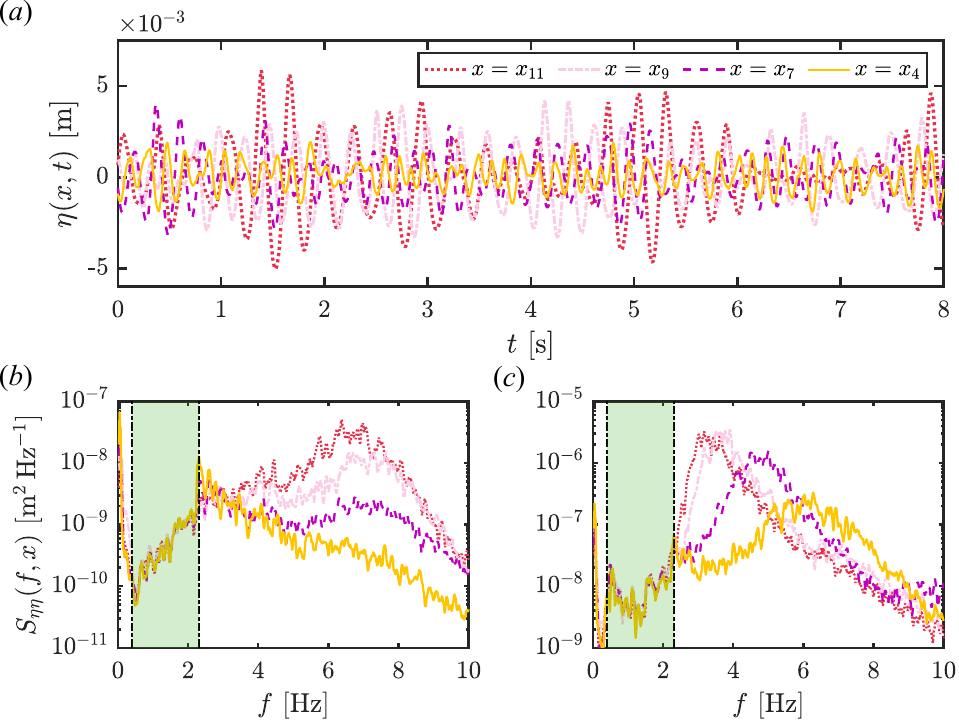}
	\caption{(\textit{a}) Surface elevation time-series of a wind-induced, statistically stationary sea state at $\overline{U_{10}}=6.0$ m$\,$s$^{-1}$, recorded at different fetches: $x_4=6$ m, $x_7=9.1$ m, $x_9=13$ m, and $x_{11}=15.7$ m (see also figure~\ref{fig: BURUS setup}\textit{a}). Also shown are the FFT variance density spectra for (\textit{b}) $\overline{U_{10}}=3.0$ m$\,$s$^{-1}$ and (\textit{c}) $\overline{U_{10}}=6.0$ m$\,$s$^{-1}$, at these fetches. Again, the two black vertical dashed lines mark the lower and upper frequency limits of the designed paddle-wave spectrum, shown as shaded regions.}
    \label{fig: BackgroundWaves fetch}
\end{figure*}

To examine the fetch dependence at these wind speeds, we plot in figure~\ref{fig: BackgroundWaves fetch}(\textit{a}) the surface elevation time-series recorded at several gauges under $\overline{U_{10}}=6.0$ m$\,$s$^{-1}$. The results show a consistent increase in wave amplitude (and scale) with fetch, and the spectra in figures~\ref{fig: BackgroundWaves fetch}(\textit{b}, \textit{c}) show how this development depends on wind speed. For both wind speeds, spectral growth occurs mainly beyond the upper frequency limit of the target paddle-wave spectrum in the higher frequency band. At $\overline{U_{10}}=3.0$ m$\,$s$^{-1}$, no distinct peak forms until the most downstream location shown, $x_{11}$, where a maximum appears at $f\approx6.8$ Hz. By contrast, at $\overline{U_{10}}=6.0$ m$\,$s$^{-1}$, a high frequency peak is present throughout; with increasing fetch, this peak both amplifies and shifts progressively towards lower frequencies.

Overall, the above observations highlight the role of overlying wind in amplifying high-frequency spectral energy in wind-induced wave fields, with stronger wind and longer fetch promoting both continued spectral growth and redistribution towards lower frequencies (longer waves). These findings agree well with previous studies of purely wind-induced sea states \citep[e.g.][]{Hara1991,Grare2013,Kumar2024,Do2024}.

\subsection{Evolution of non-breaking focused wave groups in the presence of wind forcing} \label{appA: wind-forced non-breaking waves}

Having gained insight into the behaviour of purely wind-induced wave fields, we now examine how co-flowing wind modifies weakly non-linear, non-breaking focused wave groups. To minimise the role of intrinsic non-linearity, we focus on case G2Tp13A040, which had the smallest $A$ in \textbf{BUBER} and \textbf{EURUS}. Unforced and wind-forced runs are compared at the same fetch distances, with the prescribed $x_f$ and $t_f$ kept unchanged. As mentioned earlier, the wind was allowed to blow for 8 minutes such that a statistically-steady background wave field was established before the focused wave group was generated.

\begin{figure*}
    \centering
	\includegraphics[width=0.9\columnwidth]{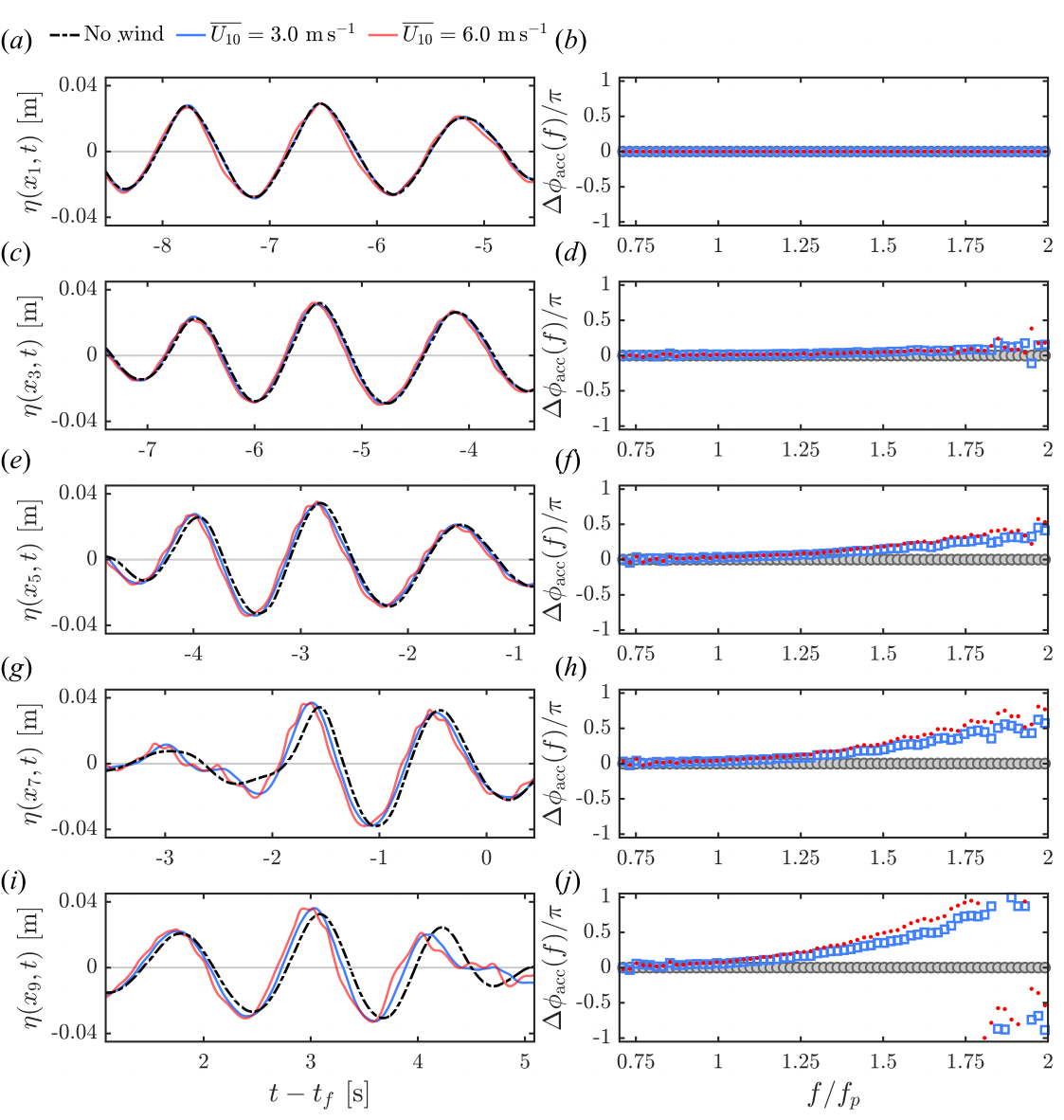}
	\caption{Surface elevation time-series $\eta(x,t)$ (left panels) and normalised accumulated phase differences $\Delta\phi_{\mathrm{acc}}(f,x)\,\pi^{-1}$ (right panels) for G2Tp13A040 from \textbf{BUBER} (no wind) and \textbf{EURUS} (with wind). The phase differences compare the phase changes from $x_1$ to each measurement location under wind and no-wind conditions, as defined in \eqref{eq: accumulated phase difference}, and are wrapped to $(-\pi,\pi]$. Records were taken at (\textit{a}, \textit{b}) $x_1=2.4$ m, (\textit{c}, \textit{d}) $x_3=4.5$ m, (\textit{e}, \textit{f}) $x_5=7.0$ m, (\textit{g}, \textit{h}) $x_7=9.1$ m, closest to the prescribed focal location, and (\textit{i}, \textit{j}) $x_9=13$ m (see also figure~\ref{fig: BURUS setup}\textit{a}). All cases had the same prescribed focal location $x_f$ and time $t_f$.}
    \label{fig: Wind Focused Eta}
\end{figure*}

The surface elevation time-series, $\eta(t)$, are shown in the left panels of figure~\ref{fig: Wind Focused Eta}, where (\textit{a, c, e,} and \textit{g}) correspond to measurements upstream of the focal point $x_f$ and (\textit{i}) is downstream. Near the wave generation paddle (\textit{a}, \textit{c}), differences between wind and no-wind conditions are very small, as the fetch is too short for substantial wind--wave interaction and the airflow itself is still developing. Further downstream (\textit{e}, \textit{g}, and \textit{i}), deviations become apparent: the wind-forced wave groups arrive earlier at a given gauge, suggesting faster propagation, with a larger shift at the higher wind speed. The crests also appear more disturbed, and slightly higher, under wind forcing.

Together, the above observations imply that the wind-forced groups no longer focus at the same time or location as the corresponding unforced groups, although the precise focal time and location were not resolved by our wave-gauge array. Indeed, \cite{Gray2021} demonstrated that under wind action, frequency components, particularly at higher frequencies, lose phase coherence at the target focal point, thereby displacing the point of maximum symmetry further downstream. Our observations are consistent with such a downstream shift under wind forcing, despite the prescribed $x_f$ remaining unchanged.

A possible contribution to this behaviour is the modification of wave dispersion by wind-induced drift currents. Under wind forcing, a near-surface drift layer develops with strong vertical shear \citep{Banner1974,Tian2013,Zou2017}. Within this layer, the wind-induced drift current has a depth-dependent velocity profile \citep{Zou2017,Chen2022}, from which Doppler-like modifications to the phase velocity of each spectral component arise. As discussed by \cite{Cummins1993}, the effect of a sheared current cannot generally be represented by simply adding the surface drift velocity to the underlying wave phase speed. For small amplitude waves on a weak current, we consider the first-order depth-weighted advection approximation of \citet{Kirby1989}, with surface tension included:

\begin{equation}
c(k)\simeq
\underbrace{\sqrt{\left(\frac{g}{k}+\frac{\sigma k}{\rho}\right)\tanh(kd)}}_{\displaystyle c_0(k)}
+
\underbrace{U_{\mathrm{eff}}(k)}_{\displaystyle \text{wind-induced advection}},
\label{eq: drift speed}
\end{equation}
where $c_0(k)$ is the gravity--capillary phase speed in the absence of a current, $\sigma$ is the surface tension, and $\rho$ is the water density. We adopt representative freshwater values of $\sigma=0.072$ N$\,$m$^{-1}$ and $\rho=1000$ kg$\,$m$^{-3}$. The effective advection velocity in \eqref{eq: drift speed} is obtained by weighting the wind-induced drift-current profile $u_{\mathrm{drift}}(z)$ over depth \citep{Kirby1989}:

\begin{equation}
U_{\mathrm{eff}}(k)=\frac{2k}{\sinh(2kd)}\int_{-d}^{0}u_{\mathrm{drift}}(z)\cosh[2k(z+d)]\,\mathrm{d}z.
\label{eq: drift effective current}
\end{equation}
Here, $u_{\mathrm{drift}}(z)$ denotes the background current velocity in the direction of wave propagation at vertical position $z$, with $z=0$ at the mean water surface and $z=-d$ at the flume bed. The dependence of $U_{\mathrm{eff}}$ on $k$ accounts for the different propagation changes experienced by different wave components (see also \S4 of \citet{Martin2026}).

At a given wavenumber $k$ the resulting phase speed change modifies the frequency relative to its current-free value, $f_0(k)=kc_0(k)/(2\pi)$, such that

\begin{equation}
f_{\mathrm{mod}}(k)=\Gamma(k)f_0(k),
\label{eq: drift f}
\end{equation}
with the modulation factor given by

\begin{equation}
\Gamma(k)=\frac{c(k)}{c_0(k)}
\simeq1+\frac{U_{\mathrm{eff}}(k)}{c_0(k)}
=1+\frac{2k}{c_0(k)\sinh(2kd)}\int_{-d}^{0}u_{\mathrm{drift}}(z)\cosh[2k(z+d)]\,\mathrm{d}z.
\label{eq: drift coefficient}
\end{equation}

To evaluate $f_{\mathrm{mod}}(k)$ and $\Gamma(k)$ from equations~\eqref{eq: drift f} and \eqref{eq: drift coefficient}, one requires the drift-current profile $u_{\mathrm{drift}}(z)$. Since neither the surface drift velocity nor its vertical profile was measured here, we adopt an exponential profile following \citet{Zou2017} and \citet{Chen2022}, expressed relative to the mean water surface:

\begin{equation}
u_{\mathrm{drift}}(z)=\left.u_{\mathrm{drift}}\right|_{z=0}e^{z/\delta^*},
\qquad -d\leq z\leq0,
\label{eq: drift profile}
\end{equation}
where $\delta^*$ is the characteristic $e$-folding thickness of the drift layer, for which the velocity decreases to $1/e$ of its surface value at $z=-\delta^*$ and attenuates to a negligible level at $z=-\pi\delta^*$ \citep{Zou2017}. We take $\delta^*=0.01$ m for both wind conditions, consistent with the exponential-profile calculations of \citet{Zou2017}. The surface drift velocity is commonly scaled with the air-side friction velocity $u_*$, and here we adopt $\left.u_{\mathrm{drift}}\right|_{z=0}=0.3u_*$, following \citet{Grare2013}. From the airflow measurements at probe 2, we obtain $u_*=0.07$ and $0.21$ m$\,$s$^{-1}$, corresponding to surface drift velocities of $0.021$ and $0.063$ m$\,$s$^{-1}$ for $\overline{U_{10}}=3.0$ and $6.0$ m$\,$s$^{-1}$, respectively. These estimates are consistent with the scaling $\left.u_{\mathrm{drift}}\right|_{z=0}=\mathcal{O}(1\%\,\overline{U_{10}})$ reported by \citet{Tian2013}, and are used here to construct representative drift profiles in equation~\eqref{eq: drift profile}. 

Using these profiles in equations~\eqref{eq: drift effective current} and \eqref{eq: drift speed} gives an initial estimate of the phase speed $c(k)$. However, to obtain $c(k)$ without truncating the current-induced correction at first order, we solve the full linear wave--current boundary value problem at each prescribed wavenumber $k$. For a steady, locally parallel, inviscid current, the vertical velocity amplitude $\widehat{w}(z)$ satisfies the Rayleigh equation \citep{Kirby1989},

\begin{equation}
[u_{\mathrm{drift}}(z)-c]
\left[
\frac{\mathrm{d}^2\widehat{w}}{\mathrm{d}z^2}
-k^2\widehat{w}(z)
\right]
-\frac{\mathrm{d}^2u_{\mathrm{drift}}}{\mathrm{d}z^2}
\widehat{w}(z)=0,
\label{eq: drift Rayleigh}
\end{equation}
subject to the bed and linearised free-surface conditions, including the restoring effects of gravity and surface tension,

\begin{equation}
\begin{aligned}
\widehat{w}(-d)&=0,\\
\left(\left.u_{\mathrm{drift}}\right|_{z=0}-c\right)^2
\left.\frac{\mathrm{d}\widehat{w}}{\mathrm{d}z}\right|_{z=0}
&-\left[
\left(\left.u_{\mathrm{drift}}\right|_{z=0}-c\right)
\left.\frac{\mathrm{d}u_{\mathrm{drift}}}{\mathrm{d}z}\right|_{z=0}
+g+\frac{\sigma k^2}{\rho}
\right]\widehat{w}(0)=0.
\end{aligned}
\label{eq: drift boundary conditions}
\end{equation}
Starting from the estimated phase speed, we integrate equation~\eqref{eq: drift Rayleigh} from the bed to the mean water surface, imposing $\widehat{w}(-d)=0$ and a fixed, non-zero value of $\mathrm{d}\widehat{w}/\mathrm{d}z$ at the bed to set the arbitrary amplitude scale. The trial phase speed $c$ is adjusted until the free surface condition in equation~\eqref{eq: drift boundary conditions} is satisfied. Repeating this procedure for each $k$ gives $c(k)$, from which we calculate $\Gamma(k)$ and $f_{\mathrm{mod}}(k)$ using equations~\eqref{eq: drift coefficient} and \eqref{eq: drift f}, respectively.

\begin{figure*}
    \centering
	\includegraphics[width=0.85\columnwidth]{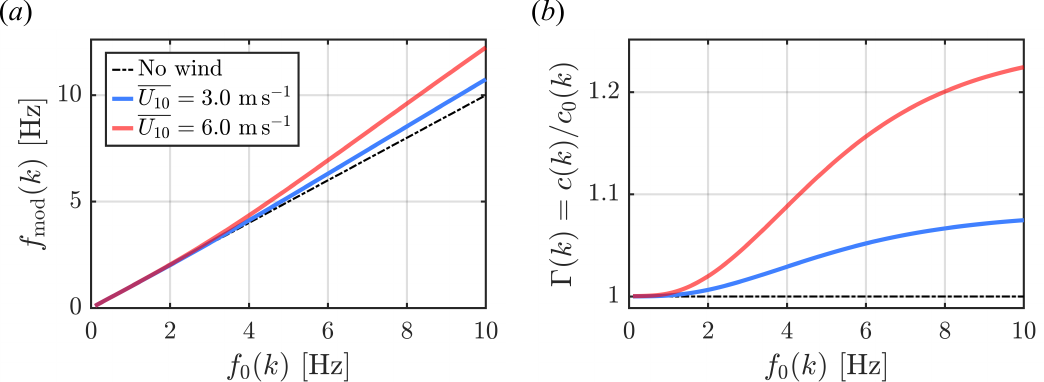}
	\caption{Wind-induced modulation of gravity--capillary wave dispersion for the exponential drift-current profiles defined in equation~\eqref{eq: drift profile}. (\textit{a}) Wind-modified frequency $f_{\mathrm{mod}}(k)$ versus the no-current frequency $f_0(k)$. (\textit{b}) Modulation factor $\Gamma(k)=c(k)/c_0(k)$ plotted against $f_0(k)$. Each comparison is made at the same wavenumber $k$.}
    \label{fig: changes in dispersion}
\end{figure*}

The results are presented in figure~\ref{fig: changes in dispersion}, where panel~(\textit{a}) shows the wind-modified frequency at a given wavenumber and panel~(\textit{b}) shows the corresponding relative change in phase speed, represented by the modulation factor $\Gamma(k)$. We confirm that the wind-induced drift current has a greater impact on higher-frequency components through more effective advection. This arises because shorter waves have wave-induced motions concentrated nearer the surface, where the drift current is strongest, whereas the motions associated with longer waves extend to greater depths over which the current rapidly weakens \citep{Martin2026}. In this sense, the finite thickness of the drift layer sets an approximate upper wavelength beyond which waves are only weakly modified \citep{Banner1974}. 

To examine this interpretation, we compare the accumulated phase differences between unforced and wind-forced wave groups in the right-hand panels (\textit{b, d, f, h,} and \textit{j}) of figure~\ref{fig: Wind Focused Eta}. These differences are referenced to the most upstream gauge at $x_1$ and defined as

\begin{equation}
\Delta\phi_{\mathrm{acc}}(f,x)
\equiv
\bigl[\phi_W(f,x)-\phi_W(f,x_1)\bigr]
-
\bigl[\phi_N(f,x)-\phi_N(f,x_1)\bigr]
\pmod{2\pi},
\label{eq: accumulated phase difference}
\end{equation}
where $\phi_W$ and $\phi_N$ are the FFT phase angles under wind-forced and unforced conditions, respectively. The phase differences are wrapped to $(-\pi,\pi]$, accounting for the apparent jumps at the high-frequency end of panel~(\textit{j}). For clarity, only the spectral energy-containing range $0.7 \leq f/f_p \leq 2$ is shown (cf. figure~\ref{fig: SIREN setup}\textit{b}). It is clear from these plots that while lower-frequency components appear visually in phase with little accumulated phase difference, positive phase offsets are identified at higher frequencies under wind forcing. These offsets generally grow with increasing fetch and wind speed.

The above phase behaviour is qualitatively consistent with the argument that higher-frequency components are more readily advected by co-flowing wind, while lower-frequency components, being intrinsically faster, are less affected. The resulting reduction in phase-speed contrast between short and long components slows their phase convergence, so that the wave group requires a longer time and distance to reach full focusing. Related changes in focusing under wind forcing have also been reported by \cite{Touboul2006} and \cite{Gray2021}.

\begin{figure*}
    \centering
	\includegraphics[width=0.9\columnwidth]{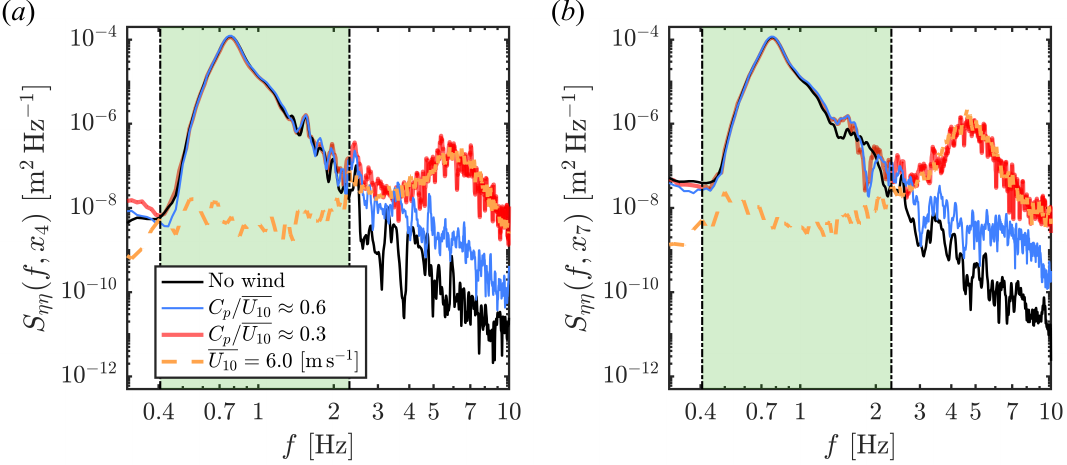}
	\caption{FFT variance density spectra for G2Tp13A040 from \textbf{BUBER} (no wind) and \textbf{EURUS} (with wind), recorded at (\textit{a}) $x_4=6.0$ m and (\textit{b}) $x_7=9.1$ m. Orange lines show spectra of purely wind-induced waves under the strongest wind speed, $\overline{U_{10}}=6.0$ m$\,$s$^{-1}$. Vertical dashed lines mark the active frequency range of wave-generation paddle, corresponding to the target spectrum. The $x$-axis is plotted on a logarithmic scale to highlight wind effects.}
    \label{fig: Wind Focused Spectra}
\end{figure*}

From an energetic perspective, we compare variance density spectra for G2Tp13A040 at two fetches and under different wind speeds in figure~\ref{fig: Wind Focused Spectra}. Within the target frequency range, wind has little influence at $x_4$ (close to the first wind probe, panel \textit{a}), but by $x_7$ changes appear at higher frequencies, about $f>1.3$ Hz (panel \textit{b}). The dominant wind effect remains beyond the designed spectral limit, where spectral energy increases substantially with wind speed. In both panels, the spectrum of wind-induced waves at $\overline{U_{10}}=6.0$ m$\,$s$^{-1}$ is also shown; its high-frequency tail accounts for most of the spectral energy in this range of the combined wind-wave spectra. This partly echoes the conclusion of \citet{Lee2020}: direct wind forcing does not necessarily strengthen higher-order non-linear wave--wave interactions, and the additional growth observed here arises primarily from (linear) wind--wave interactions.

\subsection{Summary of the wind effect on unsteady, non-breaking wave groups}\label{appA: summary}
The experimental results presented in this appendix (\ref{appA}) reveal several effects of co-flowing wind on weakly non-linear, unsteady wave groups under non-breaking conditions, consistent with the findings of \citet{Touboul2006}, \citet{Gray2021}, and \citet{Chen2022}. The main physical mechanisms are summarised as follows:
\begin{itemize}
\item The presence of overlying wind amplifies spectral energy in the higher-frequency band, introducing additional short waves whose energy is largely accounted for by the purely wind-induced components, consistent with approximately linear superposition. This appears as increased surface disturbance on the wave-group profile.

\item Wind-induced drift currents modulate wave-group dispersion, causing the groups to propagate faster and potentially focus further downstream than their equivalent wave groups when no wind is present. These effects become more pronounced at higher wind speeds.

\item The apparently contrasting observations of a forward time shift (suggesting shorter time to focus) and a downstream shift of the focal location (suggesting longer distance to focus) can be reconciled by the inconsistent influence of drift currents across spectral components. Shorter waves are more strongly advected than longer waves \citep{Banner1974,Gray2021}, reducing their relative phase-speed difference and delaying their phase alignment. As a result, the group can arrive earlier at a given gauge while still requiring a longer distance for a focused event to form.

\item Although the added spectral energy near and above the upper frequency limit of the designed spectrum is relatively small, this band is important for breaking initiation through spectral saturation \citep{Phillips1985,Kway1998}. This motivates our analysis in the main text of how wind modifies the steepness at which breaking occurs.

\end{itemize}

\section{Empirical fits for wind effects on crest geometry}
\label{appB}

Table~\ref{tab:fit_coefficients} summarises the empirical fits shown in figures~\ref{fig: steepness compare BURUS}
and~\ref{fig: leaning}, together with their coefficients,
95\% confidence intervals and $R^2$ values. These fits describe the measured relationships under the three wind conditions and facilitate comparison of the observed trends (they are not intended as general predictive laws).

\begin{table}
\centering
\small
\setlength{\tabcolsep}{3pt}
\renewcommand{\arraystretch}{1.15}

\resizebox{\textwidth}{!}{%
\begin{tabular}{cccccc}
\toprule
\multicolumn{6}{c}{
$\mathcal{S}_{\mathrm{front}}(t_b)
=\mathcal{L}_1\mathcal{S}_j(x_3)+\mathcal{L}_2$,
\quad $j=p,s,n$
} \\
\midrule
Independent variable & Figure panel & Wind condition
& $\mathcal{L}_1$ & $\mathcal{L}_2$ & $R^2$ \\
\midrule

$\mathcal{S}_p(x_3)$
& \ref{fig: steepness compare BURUS}(\textit{b})
& No wind
& $3.70\,(\pm1.25)$ & $-0.81\,(\pm0.46)$ & 0.81 \\
& & $C_p/\overline{U_{10}}\approx0.6$
& $4.25\,(\pm1.61)$ & $-1.05\,(\pm0.59)$ & 0.80 \\
& & $C_p/\overline{U_{10}}\approx0.3$
& $3.81\,(\pm1.32)$ & $-0.95\,(\pm0.49)$ & 0.83 \\[3pt]

$\mathcal{S}_s(x_3)$
& \ref{fig: steepness compare BURUS}(\textit{e})
& No wind
& $3.24\,(\pm1.07)$ & $-0.81\,(\pm0.46)$ & 0.82 \\
& & $C_p/\overline{U_{10}}\approx0.6$
& $3.45\,(\pm1.35)$ & $-0.96\,(\pm0.57)$ & 0.79 \\
& & $C_p/\overline{U_{10}}\approx0.3$
& $3.24\,(\pm1.05)$ & $-0.91\,(\pm0.45)$ & 0.84 \\[3pt]

$\mathcal{S}_n(x_3)$
& \ref{fig: steepness compare BURUS}(\textit{h})
& No wind
& $1.77\,(\pm0.55)$ & $-0.69\,(\pm0.39)$ & 0.84 \\
& & $C_p/\overline{U_{10}}\approx0.6$
& $1.91\,(\pm0.75)$ & $-0.86\,(\pm0.54)$ & 0.79 \\
& & $C_p/\overline{U_{10}}\approx0.3$
& $1.81\,(\pm0.49)$ & $-0.83\,(\pm0.35)$ & 0.88 \\
\bottomrule
\end{tabular}
}

\vspace{0.3cm}

\resizebox{\textwidth}{!}{%
\begin{tabular}{ccccccc}
\toprule
\multicolumn{7}{c}{
$\mathcal{V}
=\mathcal{E}_3+\mathcal{E}_1
\exp[-\mathcal{E}_2(\mathcal{S}_n(x_3)-0.55)]$
} \\
\midrule
Dependent variable & Figure panel & Wind condition
& $\mathcal{E}_1$ & $\mathcal{E}_2$
& $\mathcal{E}_3$ & $R^2$ \\
\midrule

$L^{\prime}(t_b)$
& \ref{fig: leaning}(\textit{a})
& No wind
& $0.33\,(\pm0.15)$ & $12.81\,(\pm10.05)$ & 0.18 & 0.72 \\
& & $C_p/\overline{U_{10}}\approx0.6$
& $0.50\,(\pm0.35)$ & $10.59\,(\pm9.10)$ & 0.18 & 0.67 \\
& & $C_p/\overline{U_{10}}\approx0.3$
& $0.54\,(\pm0.18)$ & $11.89\,(\pm4.92)$ & 0.21 & 0.88 \\[3pt]

$\zeta$
& \ref{fig: leaning}(\textit{b})
& No wind
& $0.81\,(\pm0.41)$ & $14.56\,(\pm13.35)$ & 0.35 & 0.68 \\
& & $C_p/\overline{U_{10}}\approx0.6$
& $1.28\,(\pm1.02)$ & $12.03\,(\pm11.19)$ & 0.37 & 0.66 \\
& & $C_p/\overline{U_{10}}\approx0.3$
& $1.46\,(\pm0.53)$ & $14.07\,(\pm6.04)$ & 0.44 & 0.89 \\
\bottomrule
\end{tabular}
}

\caption{Coefficients of the linear fits in figures~\ref{fig: steepness compare BURUS}(\textit{b}, \textit{e}, \textit{h}) and the exponential fits in figures~\ref{fig: leaning}(\textit{a}, \textit{b})
for the three wind conditions. The linear fits relate
$\mathcal{S}_{\mathrm{front}}(t_b)$ to the different types of upstream wave-group steepness. The exponential fits describe
$\mathcal{V}=L^{\prime}(t_b)$ or $\zeta$ as a function of
$\mathcal{S}_n(x_3)$. Parenthetical uncertainties are
half-widths of the 95\% confidence intervals.
For the exponential fits, $\mathcal{E}_3$ is fixed at the
mean of the three observations with the largest
$\mathcal{S}_n(x_3)$ in each wind condition, and the intervals
for $\mathcal{E}_1$ and $\mathcal{E}_2$ are conditional on
this fixed value. For $\mathcal{V}=L^{\prime}(t_b)$,
$\mathcal{E}_1$ and $\mathcal{E}_3$ have units of metres.
The strongest-wind linear fits are used to normalise
the data in figures~\ref{fig: steepness compare BURUS}
(\textit{c}, \textit{f}, \textit{i}).}
\label{tab:fit_coefficients}
\end{table}

\bibliographystyle{apalike}
\bibliography{Biblio}

\end{document}